\documentstyle[aps,epsfig,prl,floats]{revtex}

\newcommand{\be}{\begin{equation}}
\newcommand{\ee}{\end{equation}}
\newcommand{\bea}{\begin{eqnarray}}
\newcommand{\eea}{\end{eqnarray}}

\newcommand{\bfdelta}{ \mbox{\boldmath $\delta$} }
\newcommand{\sbfdelta}{\mbox{{\scriptsize\boldmath $\delta$}}}
\begin{document}
\draft

\title{
%Linked Cluster Series Expansions for\\
%the Spectral Weights of Many-particle States}
Dynamical Structure Factor for the Alternating Heisenberg Chain:\\
A Linked Cluster Calculation}

\author{ Chris J.~Hamer$^{(a)}$\cite{cjh},
         Weihong Zheng$^{(a)}$\cite{zwh}, and
         Rajiv R.~P.~Singh$^{(b)}$
}
\address{
$^{(a)}$ School of Physics, University of New South Wales, Sydney NSW
2052, Australia\\
$^{(b)}$ Department of Physics, University of California, Davis, CA
95616\\
}
%@twocolumnfalse\endcsname

\date{\today}
\maketitle
\widetext

\begin{abstract}
We develop a linked cluster method to calculate the
spectral weights of many-particle excitations at zero temperature.
The dynamical structure factor is expressed as a sum of `exclusive'
structure factors, each representing contributions from a given set
of excited states. A linked cluster technique to obtain high order
series expansions for these quantities is discussed. We apply these methods to
the alternating Heisenberg chain around the dimerized limit ($\lambda=0$),
where complete wavevector and frequency dependent spectral weights
for one and two-particle excitations (continuum and bound-states)
are obtained. For small to moderate values of the inter-dimer coupling
parameter $\lambda$, these lead to extremely accurate calculations
of the dynamical structure factors.
We also examine the variation
of the relative spectral weights of one and two-particle states
with bond alternation all the way up to the limit of the uniform chain
($\lambda=1$).
In agreement with Schmidt and Uhrig, we find that the spectral weight
is dominated by 2-triplet states even at $\lambda=1$,
which implies that a description in terms of triplet-pair excitations
remains a good quantitative description of the system even
for the uniform chain.
\end{abstract}

\pacs{PACS numbers:  75.40.Gb, 75.10.Jm, 75.50.Ee}
%\keywords{}

%]

%\narrowtext

\section{Introduction}

Controlled and systematic calculation of dynamical properties of
quantum lattice models remains a challenging computational task.
Despite much recent progress in developing computational
methods, such as the density matrix
renormalization group, quantum Monte Carlo and series expansions, the
dynamical properties, especially those associated with
multiparticle excitations, remain poorly understood. In many systems,
these multiparticle excitations have relatively small spectral weights.
But in low-dimensional systems, they can become extremely important
and even dominate the spectral functions. The increased frequency
and wavevector resolution of various dynamical experiments, especially
neutron scattering, necessitates going beyond the single-particle
picture and obtaining quantitatively accurate results for the full
dynamical structure factors.

One effective way to study quantum lattice models has been by
using high order power series expansions in a suitable coupling
constant. These methods have long been used to study ground state
properties and elementary excitation spectra. Recently, they
have been extended to multiparticle excitation spectra as well\cite{tre00,zhe01,kne01}.
In particular, the linked cluster expansion method\cite{gel00} is a particularly
efficient way to carry out these calculations, which provides
substantial internal checks on the validity of the expansions,
and allows one to carry out the calculations by automated computer programs.

Here, we develop a general linked-cluster formalism to calculate
the single-particle and multi-particle contributions to
the dynamical structure factor. We apply the method to the
alternating Heisenberg chain (AHC), where expansions are done around
the strong coupling limit of decoupled spin dimers up to
14th order in the ratio of coupling constants.
We calculate various properties of the multiparticle
continuum and bound states for a number of parameters including
those appropriate for the material
$Cu(NO_3)_2.2.5D_2O$, which has recently been studied by neutron scattering\cite{xu00,ten02}.

The alternating Heisenberg chain is an excellent test-bed for our present
purposes. It is a simple isotropic spin system with a finite energy gap (see Section III).
The ground state is a spin singlet, and the lowest excitations are
S = 1 triplet  states. Above this band of 1-particle states
there lies a 2-particle continuum.
It was noted recently\cite{uhr96,bou98,fle97} that
there are also 2-particle bound
states with both $S = 0$ and $S = 1$ below the continuum, as well as quintuplet
antibound states above it. Trebst et al.\cite{tre00,zhe01a} found, in fact, that over a wide
range of parameters there exist {\it two} singlet and {\it two} triplet bound
states near the Brillouin zone boundary $kd = \pi$
(where $d$ is the inter-dimer spacing). Hence the model
displays some interesting multi-particle dynamics which can be explored
both theoretically and experimentally. Neutron scattering experiments,
however, will only be sensitive to the triplet bound states, which
is what we focus on here.

%Of the two triplet bound states, the one with the larger binding energy
%is also more strongly localized. In the strong coupling limit, it
%reduces to two 1-particle triplets on neighboring sites, which are bound by the
%attractive exchange. The spectral weight for this bound state peaks near (but not exactly at)
%$k=\pi/d$.
%The second bound state is relatively less localized. In
%the strong coupling limit, the pair of triplets are separated by
%an odd-number of singlet dimers, with an amplitude which decays exponentially
%with separation. The spectral weight for this triplet state vanishes
%near $k=\pi/d$ and has a two-peaked structure as a function of
%wavevector. The spectral weight of the continuum has a sharp onset
%at the lower frequency end and goes gradually to zero at the
%upper end. It has much similarity with previous calculations
%of Uhrig and Schulz\cite{uhr96}.

There has been much discussion in the
literature\cite{den79,uhr99,sor98}
about the behaviour of the alternating chain as it approaches the uniform
limit, which is a critical point of the model. The uniform chain is gapless,
and is known to exhibit ``spinon" excitations. The alternating chain is gapped, and
is described in term of triplet excitations. The debate has concerned the
behaviour of the model near the uniform limit, and how the transition is made from the ``triplet"
mode of description to the ``spinon" mode. We discuss these questions further
in Section IV.

We also directly obtain series
expansions for sum-rules representing the total contributions
of two-particle excitations to the dynamical structure factor
summed over all wavevectors. %(at any wavevector?).
Comparison of these with the static structure
factor and the one-particle spectral weights,
and extrapolations using  approximant  methods,
leads to the conclusion that just keeping the one and two-particle
excitations leads to a highly accurate description of the
full dynamical structure factor, and this description remains
quantitatively valid even in the uniform chain limit.

The plan of this paper is as follows. First, we
discuss the formalism for calculating spectral weights using
the linked cluster method.  This is followed by detailed
calculations of the spectral functions for the alternating
Heisenberg chains. These are followed by our conclusions.

%\newpage
\section{Formalism}

We follow the formalism of Tennant et al.\cite{ten02}. The inelastic
neutron scattering cross-section\cite{mar71}

\begin{equation}
\frac{d^2\sigma}{d\Omega d\omega} \propto N \sum_{\alpha,\beta}
\frac{k_f}{k_i} |F({\bf k}) |^2(\delta_{\alpha\beta}-k_{\alpha}k_{\beta})S^{\alpha\beta}
({\bf k},\omega)
\label{eq1}
\end{equation}
is proportional to the ``dynamical response" $S^{\alpha\beta}({\bf
k},\omega)$ where ${\bf k}$ is the wavevector transfer, $F({\bf k})$ is
the magnetic form factor, N is the number of scattering centres,
 $k_i$ and $k_f$ are the momenta of
initial and final neutron states respectively, and $\alpha = x,y,z$ are Cartesian spin coordinates. The
dynamical response is the space and time Fourier transform of the
spin-spin correlation function

\begin{equation}
S^{\alpha\beta}({\bf k},\omega) = \frac{1}{2\pi N} \sum_{i,j}\int_{-\infty}^{\infty}
\exp[i(\omega t + {\bf k}\cdot ({\bf r}_i - {\bf r}_j))]\langle S^{\alpha}_j (t)
S^{\beta}_i (0)\rangle dt
\label{eq2}
\end{equation}
where $i$ and $j$ label sites of the system.
At temperature $T = 0$, the dynamical response becomes
\begin{equation}
S^{\alpha\beta}({\bf k},\omega) = \frac{1}{2\pi N} \sum_{i,j}\int_{-\infty}^{\infty}
\exp[i(\omega t + {\bf k}\cdot ({\bf r}_i - {\bf r}_j) )]\langle \psi_0 | S^{\alpha}_j (t)
S^{\beta}_i (0) | \psi_0 \rangle dt
\label{eq2a}
\end{equation}
where $| \psi_0 \rangle $ is the ground state of the Hamiltonian.
 This quantity
is referred to as the neutron scattering ``structure factor".
For the alternating
Heisenberg chain, spin conservation and isotropy in spin space
ensure that $S^{\alpha\beta} = 0$ for $ \alpha \neq \beta$, and all
diagonal spin components are equivalent,

\begin{equation}
S^{xx}({\bf k},\omega) = S^{yy}({\bf k},\omega) = S^{zz}({\bf k},\omega),
\label{eq3}
\end{equation}
and
\begin{equation}
S^{xx}({\bf k},\omega) = \frac{1}{2}S^{-+}({\bf k},\omega).
\label{eq4}
\end{equation}
Henceforth we concentrate our attention on $S^{-+}({\bf k},\omega)$.

\subsection{Integrated Structure Factor}

Integrating equation (\ref{eq2}) over energy, we get:

\begin{eqnarray}
S^{-+}({\bf k}) & = & \int_{-\infty}^{\infty} d\omega S^{-+}({\bf k},\omega) \nonumber \\
 & = & \frac{1}{N}\sum_{i,j} \exp[i{\bf k}\cdot ({\bf r}_i - {\bf r}_j)] \langle
S_j^{-}S_i^{+} \rangle_0
\label{eq5}
\end{eqnarray}
which is just the Fourier transform of the spin-spin correlation
function at $t = 0$. By translation invariance,

\begin{equation}
 \langle S_j^{-}S_i^{+} \rangle_0 = C^{-+}(i^*,  {\bf r}_i - {\bf r}_j)
\label{eq6}
\end{equation}
where $i^*$ labels the position of $i$ within the unit cell. In the present case,
the unit cell consists of a single dimer.
Therefore
\begin{equation}
S^{-+}({\bf k}) = \sum_{{\bf \sbfdelta}} C^{-+}(i^*, \bfdelta)\exp[{i{\bf k} \cdot \bfdelta}]
\label{eq7}
\end{equation}
This quantity we shall refer to as the integrated structure factor.

\subsection{Exclusive Structure Factors}

Inserting a complete set of eigenstates ${ | \psi_{\Lambda} \rangle }$ of $H$
between the spin operators in equation (\ref{eq2a}), we can express the
spin structure factor as a sum over ``exclusive" structure factors

\begin{equation}
S^{-+}({\bf k},\omega) = \sum_{\Lambda} S_{\Lambda}^{-+}({\bf k},\omega)
\label{eq8}
\end{equation}
where
\begin{equation}
S^{-+}_{\Lambda}({\bf k},\omega) = \frac{1}{2\pi N} \sum_{i,j} \int_{-
\infty}^{\infty} dt \exp[i(\omega t + {\bf k}\cdot ({\bf r}_i - {\bf r}_j))]
\langle \psi_0 | S_j^{-} (t) | \psi_{\Lambda} \rangle
\langle  \psi_{\Lambda} | S_i^{+} (0) | \psi_0 \rangle
\label{eq9}
\end{equation}
Each exclusive structure factor $ S_{\Lambda}^{-+}({\bf k},
\omega)$ gives the intensity of scattering from $| \psi_0 \rangle$ to a
specific triplet excited state $| \psi_{\Lambda} \rangle $\cite{bar99}. In the Heisenberg picture
$S_j^{-}(t)= \exp(iHt)S^-_j (0) \exp(-iHt)$ gives trivial exponentials in
$t$, so the time integral simply gives an energy-conserving delta
function
\begin{equation}
S^{-+}_{\Lambda}({\bf k},\omega) = \frac{1}{N} \delta (\omega - E_{\Lambda} + E_0)
 {\Big |} \sum_{i}
\langle \psi_{\Lambda} | S_i^{+} | \psi_0 \rangle
 \exp[i {\bf k}\cdot {\bf r}_i] {\Big |}^2
\label{eq10}
\end{equation}

Assuming the states $| \psi_{\Lambda} \rangle $ are eigenstates of
momentum, the matrix elements of the spin operators at translationally
equivalent sites are equal modulo a plane wave

\begin{equation}
\langle \psi_{\Lambda}({\bf k}) | S_i^{+} | \psi_0 \rangle
 = \langle \psi_{\Lambda}({\bf k}) | S_{i^*}^{+} | \psi_0 \rangle
 \exp[-i {\bf k}\cdot ({\bf r}_i- {\bf r}_{i^*})]
\label{eq11}
\end{equation}
Then the sum over all sites $i$ can be reduced to a sum over sites
$i^{*}$ in the unit cell
\begin{equation}
S^{-+}_{\Lambda}({\bf k},\omega) = \frac{N_c^2}{N} \delta (\omega - E_{\Lambda} + E_0)
 {\Big |} \sum_{i^*}
\langle \psi_{\Lambda} ({\bf k}) | S_{i^*}^{+} | \psi_0 \rangle
 \exp[i {\bf k}\cdot {\bf r}_{i^*}] {\Big |}^2
\label{eq12}
\end{equation}
where $N_c$ is the number of unit cells on the lattice.
It is convenient to
% Barnes et al.\cite{bar99}
define the ``reduced exclusive structure factor" as

\begin{equation}
S^{-+}_{\Lambda}({\bf k},\omega) = N_c \delta (\omega - E_{\Lambda} + E_0)
 {\Big |} \sum_{i^*}
\langle \psi_{\Lambda} ({\bf k}) | S_{i^*}^{+}  | \psi_0 \rangle
 \exp[i {\bf k}\cdot {\bf r}_{i^*}] {\Big |}^2
\label{eq13}
\end{equation}

If we sum over momenta, we obtain the autocorrelation function
\be
\Phi (\omega) = {1\over 2\pi} \sum_{{\bf k}} S^{-+} ({\bf k}, \omega)
%= {1\over 2 \pi} \int_{-\infty}^{\infty}
%e^{i \omega t} \langle \psi_0 \vert S_i^- (t) S_i^+ (0) \vert \psi_0 \rangle dt
\ee
made up of exclusive contributions
\be
\Phi_{\Lambda} (\omega) = {1\over 2\pi} \sum_{{\bf k}}  S^{-+}_{\Lambda} ({\bf k}, \omega)
\ee
For the 2-particle continuum, one can also obtain
the total auto-correlation function
$\Phi_{\Lambda} = \sum_{\omega} \Phi_{\Lambda} (\omega)$.

We turn now to a discussion of algorithms for the calculation of
exclusive structure factors within perturbation theory.
Efficient linked cluster expansion methods have long been known\cite{irv84,gel90,he90,gel00}
for calculating bulk properties of a quantum lattice system. Similar
methods for the calculation of 1-particle spectra were developed by
Gelfand\cite{gel96}, and were extended to 2-particle spectra by
Trebst et al\cite{tre00,zhe01}.

Series methods for the calculations of exclusive 1-particle
structure factors or spectral weights have been developed previously\cite{sin95}, and
have been applied  in several places
(e.g. \cite{gel89,fle97,sin99}). In this formalism, a cluster expansion was
carried out directly for the structure factor itself.
This formalism is inapplicable to the 2-particle bound states, however, because
one needs to know the wavefunction of these states beforehand.
Some leading order hand calculations for 2-particle states have recently been
made\cite{bar99,ten02}.
The method proposed here  performs a cluster expansion for
the  `exclusive matrix elements', solving the effective 2-particle
Hamiltonian to generate the required wavefunctions implicitly.
A short paper giving some of our results has appeared
recently\cite{zhe02}.
We also note at this point that Knetter {\it et al.}\cite{kne01} have used an
alternative approach based on `continuous unitary transformations' which
is also capable of giving bound state energy spectra and structure
factors to high order and in great detail, but this approach is more suitable to
low dimension case.

Let us suppose that the Hamiltonian can be decomposed
\begin{equation}
H = H_0 + \lambda V
\end{equation}
where $H_0$ is the unperturbed Hamiltonian and $V$ is to be treated as a
perturbation. We aim to expand the multiparticle dispersion relations
and structure factors in powers of the parameter $\lambda$. For
illustrative purposes, we shall use the language of the AHC, but the
formalism can be applied more generally.

\subsection{1-particle states}

At zeroth order, the `single-particle' excitations in this model consist
of triplet excitations on a single dimer which can be labelled
$ | \psi_{\Lambda} (m) \rangle$, where $m$ is the position of the excited dimer, and $
\Lambda$ labels the angular momentum eigenstate (i.e. ${\bf S}^2, S_z$,
in this case for AHC, the label $\Lambda$ must correspond to the $(S,S_z) =
(1,1)$ state). In
momentum space, the eigenstates will be

\begin{equation}
 | \psi_{ \Lambda} (k) \rangle = \frac{1}{\sqrt{N_c}} \sum_{m} \exp(i{\bf k} \cdot {\bf r}_m) |
\psi_{\Lambda} (m)  \rangle
\label{eq19}
\end{equation}
This labelling can be retained at higher orders in perturbation theory
as $\lambda$ is raised from zero.

Then the matrix element
\begin{equation}
 \langle \psi_{\Lambda} (k) | S^{+}_i | \psi_0 \rangle = \frac{1}{\sqrt{N_c}}
 \sum_m \exp(-i{\bf k} \cdot {\bf r}_m) \langle \psi_{\Lambda} (m) | S^{+}_i | \psi_0 \rangle
\label{eq20}
\end{equation}

It follows from translation invariance that the matrix element
$\langle \psi_{\Lambda} (m) | S^{+}_{i} | \psi_0 \rangle$ in this
expression is a function of $({\bf r}_{i} - {\bf r}_m)$ only, i.e., we can define the exclusive
matrix elements
\be
\Omega_{\Lambda}^{\rm 1p} (i^*, \bfdelta) \equiv
\langle \psi_{\Lambda} (m) | S^{+}_{i} | \psi_0 \rangle
\ee
where $\bfdelta$ is the distance between $i$ and $m$, $\bfdelta = {\bf r}_{i} - {\bf r}_m$,
and $i^*$ labels site $i$ within the unit cell as before.

The reduced exclusive structure factor is
\begin{equation}
% S^{-+} (Q) = {\Big |} \sum_{m,i^*,\Lambda}
% \langle x_m,\Lambda | S^{+}_{i^*} | 0 \rangle \exp[iQ(x_{i^*} - x_m)] {\Big |}^2
S^{-+}_{\Lambda} ({\bf k}) = {\Big |}
\sum_{i^*, \sbfdelta} \Omega_{\Lambda} (i^*, \bfdelta) \exp[i{\bf k}\cdot \bfdelta] {\Big |}^2
\label{eq21}
\end{equation}

\subsection{2-particle states}

Since two triplets can combine to give total spin 1, there will also be
non-zero spin structure factors for 2-particle states. The 2-particle
states can be labelled according to their unperturbed counterparts in
position space $ | \psi_{\Lambda} (m,n) \rangle$ where $m,n$ label the two
dimer positions, and $\Lambda$ the corresponding total 2-particle
 angular momentum
states. Eigenstates of the Hamiltonian can then be expanded
\begin{equation}
 | \psi_{\Lambda} ({\bf k}) \rangle = \frac{1}{\sqrt{N_c}} \sum_{m,n} f_{\Lambda}({\bf r}_m-{\bf r}_n)
 e^{i{\bf k}\cdot ({\bf r}_m+{\bf r}_n)/2} | \psi_{\Lambda} (m,n)  \rangle
\label{eq27}
\end{equation}
where $f_{\Lambda}({\bf r}_m-{\bf r}_n)$ is the 2-particle wavefunction, which only depends on the
relative distance ${\bf r}_m-{\bf r}_n$.

Similarly one can define the  2-particle
exclusive matrix elements
\be
\Omega_{\Lambda}^{\rm 2p} (i;m,n) \equiv
\langle \psi_{\Lambda} (m,n)   | S^{+}_{i} | \psi_0 \rangle
\ee

The translation invariance implies that $\Omega_{\Lambda}^{\rm 2p} (i;m,n)$
only depends on the relative distance between  $m$, $n$ and $i$, i.e.
% Translation invariance also implies that
\be
\Omega_{\Lambda}^{\rm 2p} (i;m,n)
\equiv \Omega_{\Lambda}^{\rm 2p} (i^*, {\bf r},\bfdelta)
\ee
where ${\bf r} = (2 {\bf r}_i - {\bf r}_m - {\bf r}_n)/2$ and $\bfdelta = {\bf r}_m - {\bf r}_n$.

Then, we can get the reduced 2-particle exclusive structure factor as
\bea
S^{-+}_{\Lambda}&&({\bf k},\omega) =
\delta (\omega - E_{\Lambda} + E_0)
{\Big |} \sum_{i^*, {\bf r},\sbfdelta} \Omega_{\Lambda}^{\rm 2p} (i^*, {\bf r}, \bfdelta ) f_{\Lambda}(\bfdelta)
 \exp[i{\bf k}\cdot {\bf r}] {\Big |}^2
\label{eq212p}
\eea

\subsection{Cluster expansion}

It is easy to see that the exclusive matrix elements $\Omega^{\rm 1p}_{\Lambda} (\bfdelta)$
and $\Omega^{\rm 2p}_{\Lambda} ({\bf r},\bfdelta)$
obey a simple `cluster addition' property.
If a cluster C is made up of two disconnected sub-clusters A and B, then\footnote{We suppress the starred cell index
$i^*$, henceforth.}
\bea
\Omega^{\rm 1p, C}_{\Lambda}(\bfdelta)&& = \Omega^{\rm 1p,A}_{\Lambda}(\bfdelta)
+ \Omega^{\rm 1p,B}_{\Lambda}(\bfdelta) \nonumber \\
& & \label{eq22} \\
\Omega^{\rm 2p, C}_{\Lambda}({\bf r},\bfdelta) &&= \Omega^{\rm 2p,A}_{\Lambda}({\bf r},\bfdelta)
+ \Omega^{\rm 2p,B}_{\Lambda}({\bf r},\bfdelta)
\eea
where $\Omega^{\rm 1p, A}_{\Lambda}(\bfdelta) $ is trivially {\it zero} if cluster A does
not contain both dimer $m$ and site $i$, by conservation of spin, and similarly
for $\Omega^{\rm 2p, A}_{\Lambda}({\bf r},\bfdelta)$. It
follows that the elements $\Omega_{\Lambda}$ admit a linked cluster
expansion
\begin{equation}
\Omega_{\Lambda}= \sum_{\gamma } \Omega^{\gamma}_{\Lambda}
%S^{+}_{1\Lambda}(m,i) = \sum_{\gamma \ni (m,i)}s^{\gamma,+}_{1\Lambda}(m,i)
\label{eq23}
\end{equation}
where the sum over $\gamma$ denotes a sum over all connected clusters.
Correspondingly, the perturbation series expansion for
$\Omega_{\Lambda}$ could be formulated in terms of a diagrammatic
expansion where only connected diagrams contribute, although we will not
elaborate on this approach here.

An efficient linked cluster algorithm for calculation of the structure
factors can now be formulated, following Trebst et al.\cite{tre00,zhe01}:
\begin{itemize}
\item[i)] Generate a list of connected clusters $\gamma$ appropriate to
the problem at hand (in the present case, they will simply consist of
chains of dimers of different lengths);
\item[ii)] For each cluster $\gamma$, construct matrices for the
Hamiltonian $H$ and spin operators $S^{+}_i$ in the basis of singlet and
triplet dimer states corresponding to $H_0$;
\item[iii)] `Block diagonalize' the Hamiltonian by an orthogonal
transformation
\begin{equation}
H^{\rm eff} = O^T H O
\label{eq24}
\end{equation}
as outlined by Trebst et al\cite{tre00,zhe01}, constructed order-by-order in
perturbation theory so that the 1-particle states sit in a block by
themselves, and similarly the 2-particle states, etc; with this one can compute the
exclusive  matrix elements
$ \Omega_{\Lambda} $.
%simultaneously, transform the matrices for the spin operators
%\begin{equation}
%S^{{\rm eff},+}_i = O^T S^{+}_i O
%\label{eq25}
%\end{equation}
\item[iv)] Subtract all sub-cluster contributions to get the cumulant
$ \Omega_{\Lambda} (\bfdelta)$;

\item[v)] Insert the cumulant
$ \Omega_{\Lambda} (\bfdelta)$ in Eq. (\ref{eq23}), and hence one can
build up the exclusive matrix elements for the bulk system.
% structure factors order-by-order in perturbation theory.

\item[vi)]For the 1-particle case, one can insert the exclusive matrix elements
into Eq. (\ref{eq21}) to get the series for the exclusive structure factors.
For the 2-particle case, one still needs to solve the effective 2-particle
Hamiltonian to get the wavefunctions
$f_{\Lambda}$ for the possible bound states and continuum: this can be done  by using the
finite lattice approach\cite{zhe01}. To get the series solution rather
than the numerical solution for the wavefunctions
$f_{\Lambda}$ and the 2-particle dispersion relation, one has to use
degenerate perturbation theory. With the results for exclusive  matrix elements
$ \Omega_{\Lambda} $ and wavefunctions
$f_{\Lambda}$ in hand, one can get the exclusive structure factor $S_{\Lambda}$
through Eq. (\ref{eq212p}).

\end{itemize}

\section{Results for the Alternating Heisenberg Chain}

We apply this method to investigate the spectral weights of
the  alternating Heisenberg chain, which can be described
by the following Hamiltonian
\begin{equation}
H =  \sum_{i}  {\bf S}_{2i}\cdot {\bf S}_{2i+1} +
 \lambda {\bf S}_{2i-1}\cdot {\bf S}_{2i}
\label{eqH}
\end{equation}
where the ${\bf S}_i$ are  spin-$\frac{1}{2}$ operators at site $i$,
and $\lambda$ is the alternating coupling. Here we assume that the distance
between neighboring spins are all equal and the distance
between two successive dimers is $d$,
and assume $d=1$ if it does not appear explicitly (note this is
different from our previous
paper\cite{zhe01a}, where we had taken the lattice spacing $a$ to be 1).

There is a considerable literature on this model, which has been reviewed
recently by Barnes et al.\cite{bar99}.
At $\lambda = 0$, the system consists of a chain of decoupled dimers,
and in the ground state each dimer is in a singlet
state. Excited states are made up from the three triplet excited
states on each dimer, with a finite energy gap between the singlet ground state
and the triplet excited states. This scenario is
believed\cite{duf68,bon82,jia01} to hold
right up to the uniform limit $\lambda = 1$, which corresponds to a
critical point. At $\lambda = 1$, we regain the uniform Heisenberg chain,
which is gapless.

Several theoretical
papers\cite{den79,uhr99,sor98} have discussed the
approach to the uniform limit.
% and the way in which the alternating
% system with spin-1 triplet excitations crosses over to the uniform system
% with spin-$\frac{1}{2}$ `spinon' excitations\cite{fad81}.
Analytic studies of the critical behaviour near $\lambda=1$\cite{den79}
have related the alternating chain to the 4-state Potts model, and
indicate that the ground-state energy per site $\epsilon_0(\lambda)$, and
the energy gap $\Delta(\lambda)$ should behave as
\bea
\epsilon_0 (\lambda) - \epsilon_0 (1) &\sim& \delta^{4/3}/|\ln (\delta/\delta_0)| \nonumber
\\
&& \label{eq_log} \\
\Delta (\lambda) &\sim& \delta^{2/3}/\sqrt{|\ln (\delta/\delta_0 )|} \nonumber
\eea
as $\lambda\to 1$, where $\delta=(1-\lambda)/(1+\lambda)$.
The logarithmic terms are due to the existence of a marginal variable
in the model.
% One way to view the crossover from elementary triplets to spinon excitations is as
% follows\cite{sor98}. At $\lambda=1$, the low-lying spectrum consists of a gapless
% continuum of spinon-antispinon states. For $\lambda \lesssim 1$, the
% spinon-antispinon pairs are confined by a linear potential, giving rise
% to discrete states corresponding to the triplet excitations. In the
% triplet languange, on the other hand, the triplet energy gap drops
% to zero as $\lambda\to 1$, and the multi-triplet bound states and continua
% condense to form a continuum matching the spinon-antispinon
% description\cite{sor98,zhe01a}.

Numerical studies of the model include series expansions\cite{duf68,bon82,sin99}, and
exact diagonalizations for finite lattices\cite{soo85,spr86}.
Recently, Papenbrock {\it et al.}\cite{pap02} have  carried out density-matrix
renormalization group studies on lattices up to 192 sites in extent.
They conclude that the data for the ground-state energy and
triplet energy gap are consistent with Eq. (\ref{eq_log}), but with
surprisingly large scale factors $\delta_0$ in the logarithms.
Dlog Pad\'e analysis of the series supports these conclusions\cite{sin99}.

The 2-triplet bound states were previously studied by Uhrig and
Schulz\cite{uhr96} using an RPA approach. They found a singlet bound state
below the 2-particle continuum for all momenta $k$ and over the whole
range of $1>\lambda > 0$. They also predicted a triplet bound state and a
quintuplet antibound state near $kd = \pi$ for  small $\lambda$.
These conclusions were supported in later studies
\cite{bou98,fle97,she99}.
% Trebst et al.\cite{tre00,zhe01a} found
% in a high-order series expansion study that in fact there are {\it two}
% $S = 0$ and two $S = 1$ bound states, together with two $S = 2$
% antibound states, if one goes sufficiently close to $kd = \pi$.
Barnes et al.\cite{bar99} and Tennant et al.\cite{ten02} have shown how to calculate
exclusive structure factors for the 2-particle states by low-order
series expansions in $\lambda$, and have made a comparison with experimental data for
the copper nitrate material, $Cu(NO_3)_2.2.5D_2O$. They find that the
experimental data are consistent with the existence of a bound state,
but do not yet constitute definitive proof of it. They highlight the
need for more powerful perturbative techniques to calculate these
multiparticle cross sections: the present paper is aimed at meeting this
need\cite{zhe02}.

We have recently made an extensive study of the two triplet bound states of this
model using  high-order series expansions\cite{tre00,zhe01a}. We found that in fact
there are {\it two} singlet ($S_1$ and $S_2$) and {\it two} triplet
($T_1$ and $T_2$)
bound states below the two-particle continuum, and {\it two} quintet antibound states
($Q_1$ and $Q_2$)
above the continuum, for $\lambda$ not too large.
Meanwhile, Schmidt and Uhrig\cite{sch02} have used a
different technique, the `continuous unitary transformations'
(CUTS) method, to study the model at high orders in perturbation theory.
We shall compare our results with theirs in what follows.

Here we study the structure factors using series expansions.
We have computed, up to order $\lambda^{13}$,
the series for the integrated structure factor $S(k)$ and
the exclusive structure factors for the
1-particle triplet state $S_{\rm 1p}(k)$.
For the 2-particle states, we also computed
 the total 2-particle
structure factor  $S_{\rm 2p}$
(summed over all 2-particle states) up to order $\lambda^{13}$, and to order $\lambda^{12}$ for the
exclusive  matrix elements
$ \Omega^{\rm 2p}_{\Lambda}(r,\delta) $.
With these we can compute
the structure factors for the
2-particle triplet bound states (i.e. $S_{T_1}$ for $T_1$ and $S_{T_2}$ for
$T_2$,),  and  for the 2-particle
continuum $S_{\rm 2pc}$ up to order $\lambda^{14}$.
Integrating the structure factor over momentum $k$,
we can also compute the series for
the auto-correlation function defined in Eq. (15). We have computed the
auto-correlation functions
for the 1-particle ($\Phi_{\rm 1p}$), and
2-particle sectors ($\Phi_{\rm 2p}$), and the individual
auto-correlation functions for
2-particle bound states $T_1$ ($\Phi_{T_1}$)
and $T_2$ ($\Phi_{T_2}$) and the 2-particle continuum ($\Phi_{\rm 2pc}$).
Full series for $S(k)$ and $S_{\rm 1p}(k)$ are given in Tables
\ref{tab1} and \ref{tab2}.
Some other selective series (at $kd\to 0$, $kd=\pi$, and $2\pi$)
 are given in Table \ref{tab4}, other series are available on request.

The 1-particle structure factor has been computed to
order $\lambda^3$ by Barnes {\it et al.}\cite{bar99}, but
our series disagree with their results from second order.
%They assumed a form for the structure factor which is only
% valid in leading order.
In their calculation, they neglect the first term
in their Eq. (56);
they claim this term is zero due to a symmetry relation,
but actually this only holds in leading order.
Recently
M\"uller and Mikeska\cite{mul02} have extended the series for the
1-particle structure factor $S_{\rm 1p}(k)$ to order $\lambda^{10}$,
and our series agree with theirs.
The auto-correlation functions
for 1-particle ($\Phi_{\rm 1p}$) and
2-particle ($\Phi_{\rm 2p}$) states have been recently computed up to order $\lambda^7$
by Schmidt and Uhrig\cite{sch02}, and our results also agree with theirs,
but extend the series by up to 6 terms.
As a byproduct of our calculation, we have computed the
series for 1-triplet excitation spectra up to order $\lambda^{13}$, and
the series for 2-triplet excitation spectra up to order $\lambda^{12}$,
this extends the previous calculations\cite{zhe01a} by two terms for 1-triplet
excitation spectra, and by 1 term for 2-triplet excitation spectra.
These series are available on request.
We use  Dlog Pad\'e approximants and  integrated differential
approximants\cite{gut} to obtain numerical results up to $\lambda=1$.

\subsection{The integrated structure factor $S$}

The integrated structure factor\footnote{We drop the superscripts $-+$ henceforth.}
% $S$  is defined by
% \be
% S(k) = \frac{1}{N} \sum_{i,j} \langle  \psi_0 \vert (S_i^{+} S_j^{-} + S_i^{-} S_j^{+}) \psi_0\rangle
% \exp [{i {\bf k} \cdot ({\bf r}_i - {\bf r}_j)}]
% \ee
$S$ has been computed up to order $\lambda^{13}$. It can be expressed as
\be
S(kd) = 1 - \cos(kd/2) + \sum_{m=1}^{\infty} \sum_{n=1}^{2m+1}
 a_{n,m} \lambda^m \cos (n kd/2)
\ee
 The  series coefficients $a_{n,m}$
are given in Table \ref{tab1}. Note that there are
two general relations for these coefficients:
\bea
&&\sum_{n=1}^{2m+1} a_{n,m} =0 \nonumber \\
&& \label{S_relation} \\
&&  a_{2m,m}=-2 a_{2m+1,m} \nonumber
\eea
With this, one can easily prove that $S(k)$ can be written in the following form with a
common factor ${\sin^2 (kd/4)}$
at all orders
\be
S(kd) = \sin^2 (kd/4) \sum_{m=0}^{\infty} \sum_{n=0}^{2m}
 b_{n,m} \lambda^m \cos (n kd/2)
\ee
where one of the coefficients $b_{2m-1,m}$ happens to be zero always for any $m$, due to
the second relation in Eq. (\ref{S_relation}).
For example, the series up to order $\lambda^4$ can be written as
\bea
S(kd) &=& {\sin^2 (\frac{kd}{4})} {\Big [}
2 + \lambda\,\cos (kd) +
  \frac{{\lambda}^2 }{8}
  \left( 2 + 4\,\cos (\frac{kd}{2}) + 3\,\cos (kd) + 3\,\cos (2\,kd) \right)
      \nonumber \\
&&   + \frac{{\lambda}^3 }{96}
     \left( 12 + 24\,\cos (\frac{kd}{2}) + 13\,\cos (kd) + 16\,\cos (\frac{3\,kd}{2}) +
       28\,\cos (2\,kd) + 15\,\cos (3\,kd) \right)  \nonumber \\
&&      +
  \frac{{\lambda}^4 }{4608}
  {\Big (} 346 + 692\,\cos (\frac{kd}{2}) + 669\,\cos (kd) +
       912\,\cos (\frac{3\,kd}{2}) + 784\,\cos (2\,kd) + 304\,\cos (\frac{5\,kd}{2}) \nonumber \\
&&  +  852\,\cos (3\,kd) + 315\,\cos (4\,kd) {\Big )}
       + O(\lambda^5) {\Big ]}  \label{eq_S}
%
%   1 - \cos ({\frac{k}{2}}) + {\frac{\lambda}{4}}
%        \left( -\cos ({\frac{k}{2}}) + 2\,\cos (k) -
%          \cos ({\frac{3\,k}{2}}) \right)
%          +
%    {\frac{{\lambda^2}}{32}} \left( \cos ({\frac{k}{2}}) + 2\,\cos (k) -
%          6\,\cos ({\frac{3\,k}{2}}) + 6\,\cos (2\,k) -
%          3\,\cos ({\frac{5\,k}{2}}) \right)
% \nonumber \\
% &&         +
%    {\frac{{\lambda^3} }{384}}
%    \left( 11\,\cos ({\frac{k}{2}}) - 14\,\cos (k) -
%          9\,\cos ({\frac{3\,k}{2}}) + 40\,\cos (2\,k) -
%          43\,\cos ({\frac{5\,k}{2}}) + 30\,\cos (3\,k) -
%          15\,\cos ({\frac{7\,k}{2}}) \right)
% \nonumber \\
% &&         +
%    {\frac{{\lambda^4}}{18432}}
%    {\Big (} 23\,\cos ({\frac{k}{2}}) - 266\,\cos (k) +
%          371\,\cos ({\frac{3\,k}{2}}) + 352\,\cos (2\,k) -
%          1028\,\cos ({\frac{5\,k}{2}}) + 1400\,\cos (3\,k) \nonumber \\
% &&          -
%          1167\,\cos ({\frac{7\,k}{2}}) + 630\,\cos (4\,k) -
%          315\,\cos ({\frac{9\,k}{2}}) {\Big )} + O(\lambda^5)
\eea

\begin{figure}
\begin{center}
% \vskip -7mm
 \epsfig{file=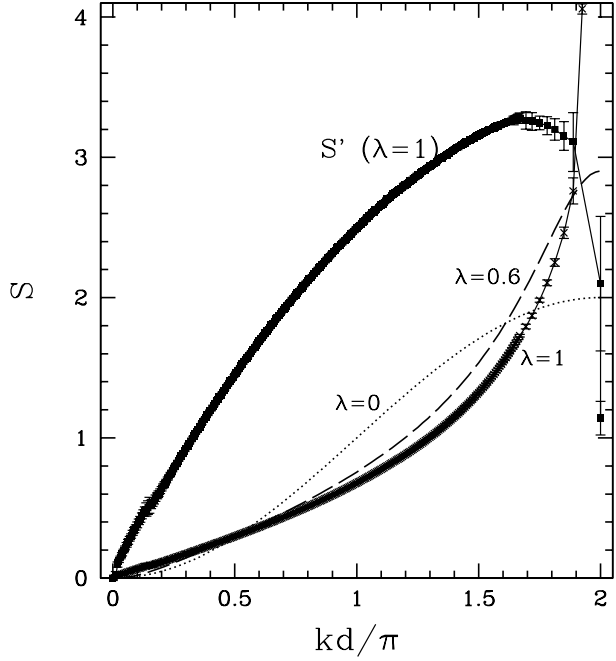,width=9cm}
  \vskip 5mm
 \caption[]
         {The integrated structure factor $S$ versus $k$ for
         $\lambda=0$ (dotted line), 0.6 (dashed line),  1 (crosses).
         Also shown is the quantity
          $S'\equiv 6 S [- 2\pi\ln(1-\frac{k}{2\pi})/k]^{-3/2}$
           for $\lambda=1$ (squares).}
 \label{fig_totS_k}
 \end{center}
\end{figure}

To analyze the series $S(kd)$,
we firstly consider $S(kd)$ near two special momentum points, $kd=0$ and $kd=2\pi$.
We know that $S=0$ at  $k=0$, but it is interesting to study how it vanishes as $k\to 0$.
For small $\lambda$, it is trivial to see from Eq.(\ref{eq_S}) that
$S\propto k^2$ as $k\to 0$. Thus we can define
$R=\lim_{kd\to 0} S(kd)/(kd)^2$, and the series for $R$ is given in Table \ref{tab4}.
Applying Dlog Pad\'e approximants to this series, we find that $R$ diverges at $\lambda=1$,
with a critical index
about -0.71\footnote{Within the accuracy of our calculations this is
consistent with a value of 2/3 with logarithmic corrections}. This implies that for the uniform chain, $\lambda=1$,
$S(kd)$ no longer vanishes quadratically with $k$. Naively one might then
expect it to vanish linearly as $k\to 0$:
some evidence for this is shown in Fig. \ref{fig_totS_k}.

For the uniform chain $\lambda=1$, and near $kd\to 2\pi$, Affleck has argued\cite{aff98}
the asymptotic form for $S(kd)$ as
\be
S(kd) = \frac{8}{3 (2\pi)^{3/2}} |\ln (\pi - kd/2) |^{3/2}
\ee

This implies that for $kd=2\pi$ and  as $\lambda\to 1$,
the asymptotic form for
$S(2\pi)$ diverges as
\be
S(2\pi) \propto [- \ln (1-\lambda)]^{3/2} \quad \lambda \to 1 \label{C_S}
\ee
If so, one should be able to see a critical point at $\lambda=1$
with critical index -1 if one applies Dlog Pad\'e approximants\cite{gut} to the
series $\partial S^{2/3}/\partial \lambda$.  This is indeed the case as we can
see from Table \ref{tab5}.
Note, however, that very similar results are obtained for any index
between 3/2 and 1, so this is not a very sensitive test.
% (Note that this do not look better than
% using the form $S(2\pi) \propto  \ln (1-\lambda)$,
% the best one with both the critical point closer to 1 and the critical
% index closer to -1 is probably
% the form  $S(2\pi) \propto [- \ln (1-\lambda)]^{4/3}$ or
% $S(2\pi) \propto [- \ln (1-\lambda)]^{5/4}$, see Table \ref{tab5}).

Assuming the above asymptotic form for  $S(2\pi )$ is correct,
we can also estimate the prefactor of this  asymptotic form by using  integrated
different approximants\cite{gut} to extrapolate the series for
$S(2\pi)/[1 - \ln(1 - \lambda)]^{3/2}$ to $\lambda=1$. The result is:
\be
S(2\pi ) = 0.19(2) [-\ln (1-\lambda ) ]^{3/2} \quad \lambda\to 1
\ee
The prefactor agrees with Affleck's prediction  of $8/3(2\pi)^{3/2}=0.16932$.

For $0<kd<2\pi$, one expects $S$ to be finite for any $\lambda$.
The results for $S$ versus momentum $k$ for
$\lambda=0$, 0.6, and 1 are shown in Fig. \ref{fig_totS_k}.
Note that $\int_0^{2\pi} S(k) dk= 2 \pi$ (here we set $d=1$), independent of $\lambda$, so
the area under each curve is the same.
Also shown in the figure are the results for
$S'\equiv 6 S [- 2\pi\ln(1-\frac{k}{2\pi})/k]^{-3/2}$ at
$\lambda=1$.
%We can see that $S'$ is close to a straight line,
%i.e. $S(k)$ is close to the form $-\ln (1-\frac{k}{2\pi})$, which also
%satisfies the condition $\int_0^{2\pi} S(k) dk= 2 \pi$.

\begin{figure}
\begin{center}
% \vskip -7mm
 \epsfig{file=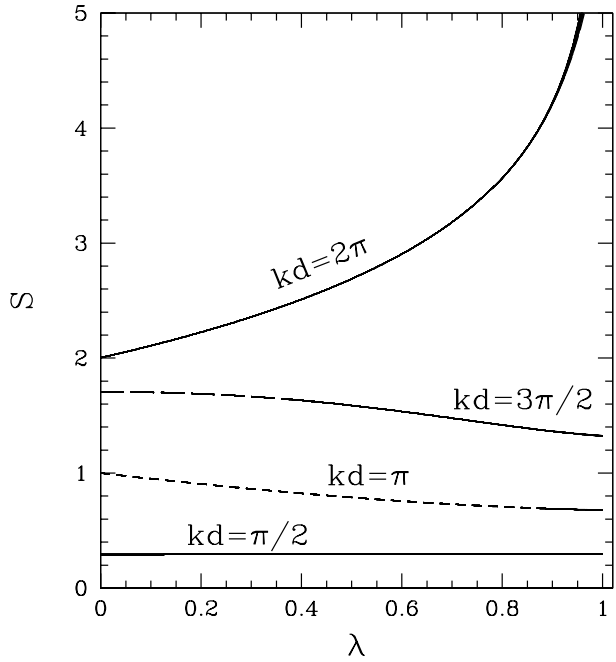,width=9cm}
  \vskip 5mm
 \caption[]
         {The integrated structure factor $S$ versus $\lambda$ for
         $kd=\pi/2$, $\pi$, $3\pi/2$ and $2\pi$.
         For each value of $kd$, about 20 different integrated differential
approximants to the high-temperature series are shown,
though most of them are indistinguishable
on the scale of this figure.}
 \label{fig_totS_x}
 \end{center}
\end{figure}

For fixed values of $k$, Fig. \ref{fig_totS_x} shows the
integrated structure factor $S$ versus $\lambda$, where for each value
of $k$, about 20 different integrated differential approximants
to the series are shown. We can see that the results converge
very well out to $\lambda=1$. The logarithmic divergence as $\lambda \to 1$
for the case $kd=2\pi$ is clearly evident.

\subsection{The 1-particle spectral weight}

The exclusive 1-particle structure factor $S_{\rm 1p}(kd)$
has been computed up to order $\lambda^{13}$.
$S_{\rm 1p}$ can also be expressed as
\be
S_{\rm 1p}(kd) = \sum_{n,m} a_{n,m} \lambda^m \cos (n kd/2)
\ee
 The  series coefficients $a_{n,m}$
are given in Table \ref{tab2}.
The coefficients $a_{n,m}$ also satisfy Eq. (\ref{S_relation}), and
$S_{\rm 1p}(kd)$
also has a common factor ${\sin^2 (kd/4)}$
for all orders. The series up to order $\lambda^4$ can be written as
\bea
S_{\rm 1p} (kd) &=& {\sin^2 (\frac{kd}{4})} {\Big [}
2 + \lambda\,\cos (kd)
+   \frac{{\lambda}^2\,}{8}
  \left( -1 + 8\,\cos (\frac{kd}{2}) + 2\,\cos (kd) + 3\,\cos (2\,kd) \right)
        \nonumber \\
&&     + \frac{{\lambda}^3 }{96}
     \left( 36\,\cos (\frac{kd}{2}) + 17\,\cos (kd) + 8\,\cos (\frac{3\,kd}{2}) + 32\,\cos (2\,kd) +
       15\,\cos (3\,kd) \right)  \nonumber \\
&&       +
  \frac{{\lambda}^4 }{4608}
  {\Big(} 342 + 384\,\cos (\frac{kd}{2}) + 784\,\cos (kd) +
       856\,\cos (\frac{3\,kd}{2}) + 675\,\cos (2\,kd) + 368\,\cos (\frac{5\,kd}{2}) \nonumber \\
&&    +  804\,\cos (3\,kd) + 315\,\cos (4\,kd) {\Big)}
       + O(\lambda^5) {\Big ]}
%
%   1 - \cos ({\frac{k}{2}}) +
%   {\frac{\lambda }{4}}
%        \left( -\cos ({\frac{k}{2}}) + 2\,\cos (k) -
%          \cos ({\frac{3\,k}{2}}) \right)
%          +
%    {\frac{{{\lambda}^2}}{32}}
%    {\Big (} -10 + 16\,\cos ({\frac{k}{2}}) -
%          4\,\cos (k) - 5\,\cos ({\frac{3\,k}{2}})  \nonumber \\
% &&   + 6\,\cos (2\,k) - 3\,\cos ({\frac{5\,k}{2}}) {\Big )}
%          + {\frac{{{\lambda}^3}}{384}}
%    {\Big(} -36 + 55\,\cos ({\frac{k}{2}}) -
%          10\,\cos (k) - 33\,\cos ({\frac{3\,k}{2}}) + 56\,\cos (2\,k) -
%          47\,\cos ({\frac{5\,k}{2}}) \nonumber \\
% &&         + 30\,\cos (3\,k) -
%          15\,\cos ({\frac{7\,k}{2}}) {\Big )}
%        + {\frac{{{\lambda}^4} }{18432}}
%    {\Big (} 300 - 700\,\cos ({\frac{k}{2}}) +
%          328\,\cos (k) + 253\,\cos ({\frac{3\,k}{2}}) + 126\,\cos (2\,k) \nonumber \\
% &&         -
%          743\,\cos ({\frac{5\,k}{2}}) + 1240\,\cos (3\,k) -
%          1119\,\cos ({\frac{7\,k}{2}}) + 630\,\cos (4\,k) -
%          315\,\cos ({\frac{9\,k}{2}}) {\Big )} + O(\lambda^5)
\eea

Let us  analyze the behaviour of $S_{\rm 1p}$ when $\lambda\to 1$.
For $kd=2\pi$, our analysis shows that $S_{\rm 1p}$ is finite as $\lambda\to 1$, so
$S_{\rm 1p}/S$  vanishes like $1/[-\ln(1-\lambda)]^{3/2}$.
For the case  $kd=0$, we again define
$R_{\rm 1p}\equiv \lim_{kd\to 0} S_{\rm 1p}(kd)/(kd)^2$ (note that $R_{\rm 1p}$ differs from
$R$ from order $\lambda^4$).  The Dlog Pad\'e approximants to $R_{\rm 1p}$
show that it diverges at $\lambda=1$
with critical index about -0.71. This again implies that for the
uniform chain case ($\lambda=1$),
$S_{\rm 1p}$ no longer vanishes as $k^2$.
Now let us analyze the series for the relative spectral weight
$R_{\rm 1p}/R$.
% the series for $R_{\rm 1p}/R$ is
% \bea
% \frac{R_{\rm 1p}}{R} &=& 1 - \frac{173\,\lambda^4}{4608} +
%   \frac{2657\,\lambda^5}{73728} + 0.0108813\,\lambda^6 -
%   0.0210969\,\lambda^7 - 0.0144688\,\lambda^8 \nonumber \\
% &&  +
%   0.0366015\,\lambda^9 - 0.0917641\,\lambda^{10} -
%   0.0291874\,\lambda^{11} + 0.0295493\,\lambda^{12} + O(\lambda^{13})
% \eea
The Dlog Pad\'e approximants to the series
$R_{\rm 1p}/R$ show no singularity near $\lambda=1$. This means that
as $\lambda\to 1$, $R_{\rm 1p}/R$ remains finite.
Our extrapolations show $R_{\rm 1p}/R=0.993(1)$
at $\lambda=1$.
% Since we expect
% $S_{\rm 1p}/S$ become zero at $\lambda=1$ for all $k$, there are probably
% a discontinue between  $\lim_{\lambda\to 1} R_{\rm 1p}/R$ and $\lim_{k\to 0}S_{\rm 1p}/S$
% at $\lambda=1$, due to the appearance of $k$ term at $\lambda=1$ for
% $S(k)$ at small $k$ limit.

\begin{figure}
\begin{center}
% \vskip -7mm
 \epsfig{file=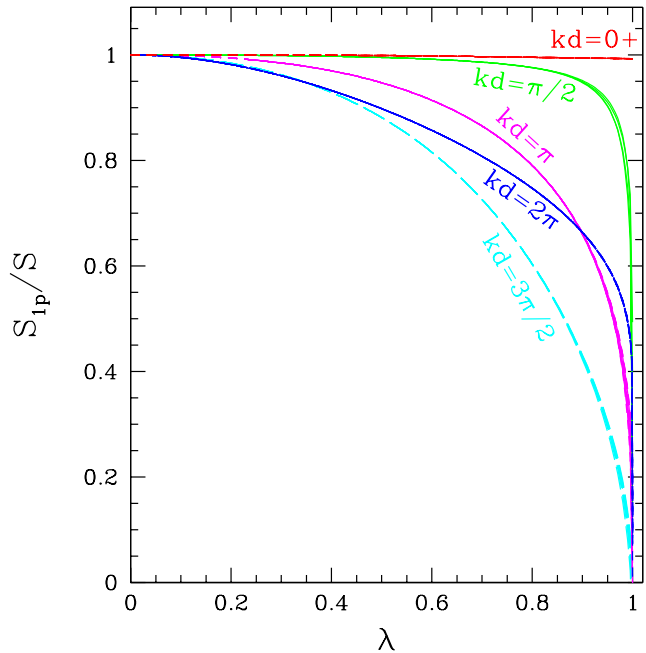,width=9cm}
  \vskip 5mm
 \caption[]
         {The relative 1-particle weight $S_{\rm 1p}/S$ versus $\lambda$ for
         $kd=0+$, $\pi/2$, $\pi$, $3\pi/2$ and $2\pi$.
         Several different integrated differential approximants to
the series are shown.
% Also shown as red lines are the results for $R_{\rm 1p}/R$ and $50(R_{\rm 1p}/R-0.99)$.
}
 \label{fig_S1p_o_S_x}
 \end{center}
\end{figure}

For $0<kd<2\pi$, the analysis of the series $S_{\rm 1p}$ by the
Dlog Pad\'e approximants  shows that it vanishes with a behavior close to
 $(1-\lambda)^{1/3}$. Since $S$ remains finite, we thus expect
that $S_{\rm 1p}/S$ vanishes like $(1-\lambda)^{1/3}$.
This agrees with the analysis of Schmidt and Uhrig\cite{sch02}, who
argued that the 1-particle spectral weight should vanish like
$\sqrt{\Delta}$, i.e.  like $\delta^{1/3}/|\ln (\delta/\delta_0)|^{1/4}$,
where $\delta =(1-\lambda)/(1+\lambda)$.
Fig. \ref{fig_S1p_o_S_x} shows the
relative 1-particle weight $S_{\rm 1p}/S$ versus $\lambda$
at selected values of $kd$.
% Values are also shown for $R_{\rm 1p}/R$ and $50(R_{\rm 1p}/R-0.99)$ versus $\lambda$.
It can be seen that for any non-zero value of
$k$, $S_{\rm 1p}/S$ decreases abruptly to zero as $\lambda\to 1$. Only at  $kd= 0+$,
does $S_{\rm 1p}/S$ remain finite (about 0.993) in the limit $\lambda=1$;
but by then $S$ has itself decreased to zero.
% the decrease towards zero becomes  sharper and sharper, until
%at $kd=0+$, $S_{\rm 1p}/S$ remains finite (about 0.993) in the limit $\lambda=1$.

The overall picture that emerges is that as $\lambda\to 1$
the triplet energy gap goes to zero, and the spectral weight associated
with the 1-triplet state also vanishes\cite{sin99}. This would seem to agree with the
idea that the ``spinons", rather than the triplet states, act as the
elementary excitations in the model in the uniform limit $\lambda\to 1$.

% Overall, then, the picture that emerges is that as $\lambda\to 1$
% the energy gap goes to zero, but the spectral weight associated with
% the triplets also vanishes, so that they no longer act as the
% elementary excitations in the model.

\begin{figure}
\begin{center}
% \vskip -7mm
 \epsfig{file=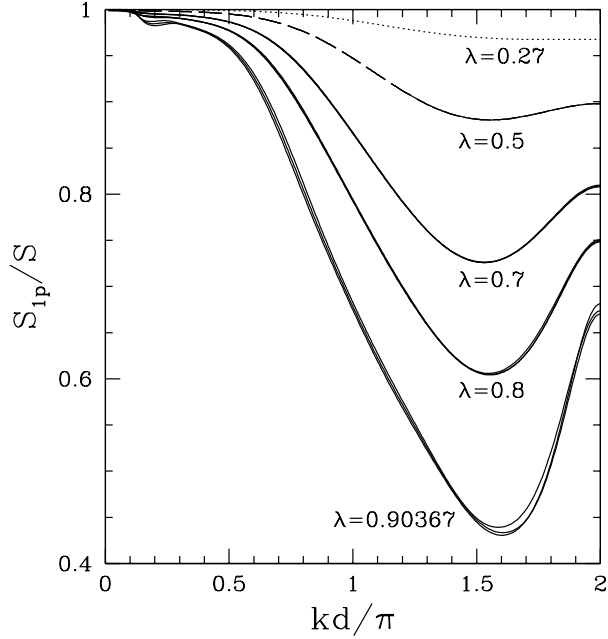,width=9cm}
  \vskip 5mm
 \caption[]
         {The relative 1-particle weight $S_{\rm 1p}/S$ versus $k$ for
         $\lambda=0.27$,  0.5,  0.7, 0.8, 0.90367.
         The results of the three highest orders are plotted.}
 \label{fig_S_o_S1p_k}
 \end{center}
\end{figure}

Fig. \ref{fig_S_o_S1p_k} shows the relative 1-particle weight $S_{\rm 1p}/S$
versus $k$ for $\lambda=0.27$,  0.5,  0.7, 0.8, 0.90367,
where $\lambda=0.90367$ is the estimated coupling for CuGeO$_3$\cite{uhr96}.
For $\lambda=0.90367$, we can see that $S_{\rm 1p}/S$ is about 1 for small $k$, and
has a minimum
around $kd = 1.6\pi$ with value 0.43, whereas Uhrig and Schulz\cite{uhr96}
estimated $S_{\rm 1p}/S=0.28$.

\subsection{The 2-particle spectral weight}
As we mentioned before, there are two triplet bound states ($T_1$ and $T_2$)
in this system. Before we discuss the spectral weight for individual states, let us first discuss the total spectral weight (the sum rule)
for all 2-particle states.

\subsubsection{The total 2-particle spectral weight}
The total 2-particle structure factor ($S_{\rm 2p}$)
has been computed up to order $\lambda^{12}$. This quantity
$S_{\rm 2p}(kd)$ also has a common factor ${\sin^2 (kd/4 )}$
for all orders, and the series up to order $\lambda^4$ can be written as
\bea
S_{\rm 2p}(kd) &=&  {\sin^2 (\frac{kd}{4})} {\Big [}
  {\lambda}^2\,{\sin^4 (\frac{kd}{4})}
 +  \frac{{\lambda}^3}{12}
 \left( 3 + 2\,\cos (\frac{3\,kd}{2}) \right) \,{\sin^2 (\frac{kd}{4})}
 + \frac{{\lambda}^4 }{4608}
 {\Big (} -57 + 372\,\cos (\frac{kd}{2}) - 143\,\cos (kd) \nonumber \\
&& +  64\,\cos (\frac{3\,kd}{2}) + 108\,\cos (2\,kd)
  - 64\,\cos (\frac{5\,kd}{2}) + 48\,\cos (3\,kd)
       {\Big )}  + O (\lambda^5) {\Big ]}
%   {\frac{{{\lambda}^2} }{32}}
%   \left( 10 - 15\,\cos ({\frac{k}{2}}) + 6\,\cos (k) -
%          \cos ({\frac{3\,k}{2}}) \right)
%                   +
%    {\frac{{{\lambda}^3} }{96}}
%    \left( 9 - 11\,\cos ({\frac{k}{2}}) - \cos (k) +
%          6\,\cos ({\frac{3\,k}{2}}) - 4\,\cos (2\,k) + \cos ({\frac{5\,k}{2}})
%           \right)  \nonumber \\
% &&          + {\frac{{{\lambda}^4} }{18432}}
%        \left( -486 + 1001\,\cos ({\frac{k}{2}}) - 722\,\cos (k) +
%          163\,\cos ({\frac{3\,k}{2}}) + 216\,\cos (2\,k) -
%          284\,\cos ({\frac{5\,k}{2}}) + 160\,\cos (3\,k) -
%          48\,\cos ({\frac{7\,k}{2}}) \right)
\eea

Here $S_{\rm 2p}\propto k^2$ in the limit $k\to 0$, and so we again
define $R_{\rm 2p}\equiv \lim_{k\to 0} S_{\rm 2p}(kd)/(kd)^2$. Note that
$R_{\rm 2p}$ is nonzero from order $\lambda^4$, while
$S_{\rm 2p}$ with $k\neq 0$ is nonzero from order $\lambda^2$.
%The series for $R_{\rm 2p}$ is also given in Table \ref{tab4}.

\begin{figure}
\begin{center}
% \vskip -7mm
 \epsfig{file=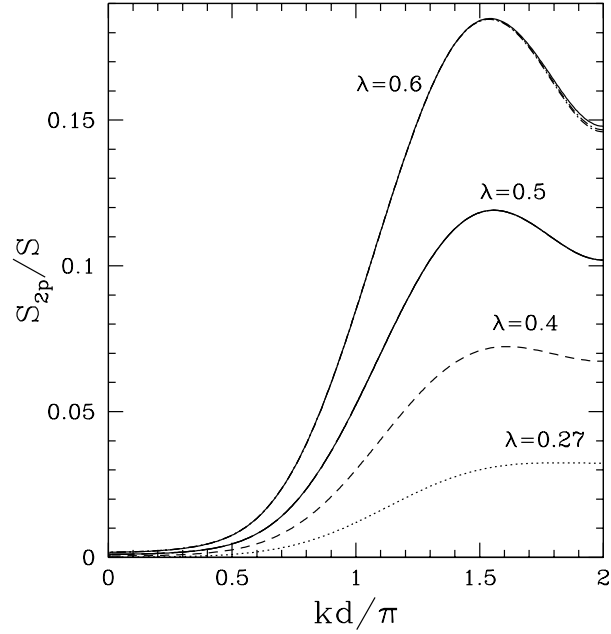,width=9cm}
  \vskip 5mm
 \caption[]
         {The relative 2-particle weight $S_{\rm 2p}/S$ versus $k$ for
         $\lambda=0.27$,  0.4, 0.5,  0.6.
         The results of the three highest orders are plotted.}
 \label{fig_S2p_o_S_k}
 \end{center}
\end{figure}

Fig. \ref{fig_S2p_o_S_k} shows the total 2-particle weight $S_{\rm 2p}/S$
versus $k$ for various $\lambda$.
For large $\lambda$, $S_{\rm 2p}/S$ has a maximum around
$kd=1.6\pi$. For $\lambda=0.6$, it has a value of about  0.1848
at its maximum, while the value for $k=0$ is only 0.00177, about 100 times
smaller than the maximum.

\begin{figure}
\begin{center}
% \vskip -7mm
 \epsfig{file=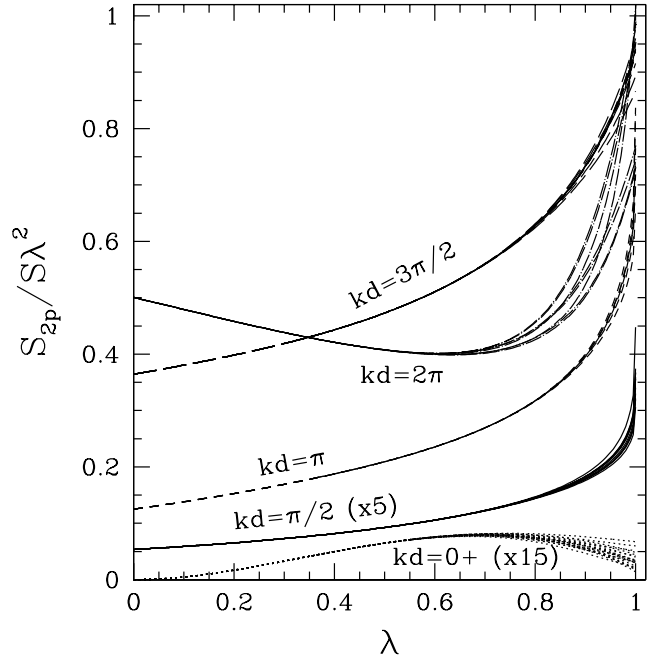,width=9cm}
  \vskip 5mm
 \caption[]
         {The relative 2-particle weight $S_{\rm 2p}/S$ versus $\lambda$ for
         $kd=0+$, $\pi/2$, $\pi$, $3\pi/2$ and $2\pi$. The results for
         $kd = 0+$ ($\pi/2$) are multiplied by a factor 15 (5) to make
         them more visible on the graph.
         Several different integrated differential approximants to
the series are shown.
%Also shown as red lines are the results for $15R_{\rm 2p}/(\lambda^4 R)$ at $kd=0$.
}
 \label{fig_S2p_o_S_x}
 \end{center}
\end{figure}

For fixed values of $k$, Fig. \ref{fig_S2p_o_S_x} shows the
relative total 2-particle weight $S_{\rm 2p}/S$ versus $\lambda$
 for $kd=\pi/2$, $\pi$, $3\pi/2$ and $2\pi$.
We can see that there are some sharp increases near $\lambda=1$
which make it difficult to estimate the results;
nevertheless, we estimate, at $\lambda= 1$,
the 2-particle states have about 90\%
of the total weight at $kd=3\pi/2$, about 80\%
of the weight at $kd=2\pi$,
about 65\%
of the weight at $kd=\pi$, and only about
10\% of the weight at $kd=\pi/2$.
We also show the results for $15S_{\rm 2p}/(\lambda^2 S)$ in the special case $k=0$,
% which decreases with $\lambda$,
%(after consider the factor
%of $\lambda^4$, $R_{\rm 2p}/R$
which show quite different behaviour from other $k$ in this figure: as $\lambda$ increases, it
increases firstly, then decreases,
so that the relative weight at $\lambda=1$ is tiny (less than 0.5\%).
This agrees with the fact that  the relative weight
for the 1-particle state, $R_{\rm 1p}/R$, is almost 99.27\% at $\lambda=1$
for this value of $k$.

\begin{figure}
\begin{center}
% \vskip -7mm
 \epsfig{file=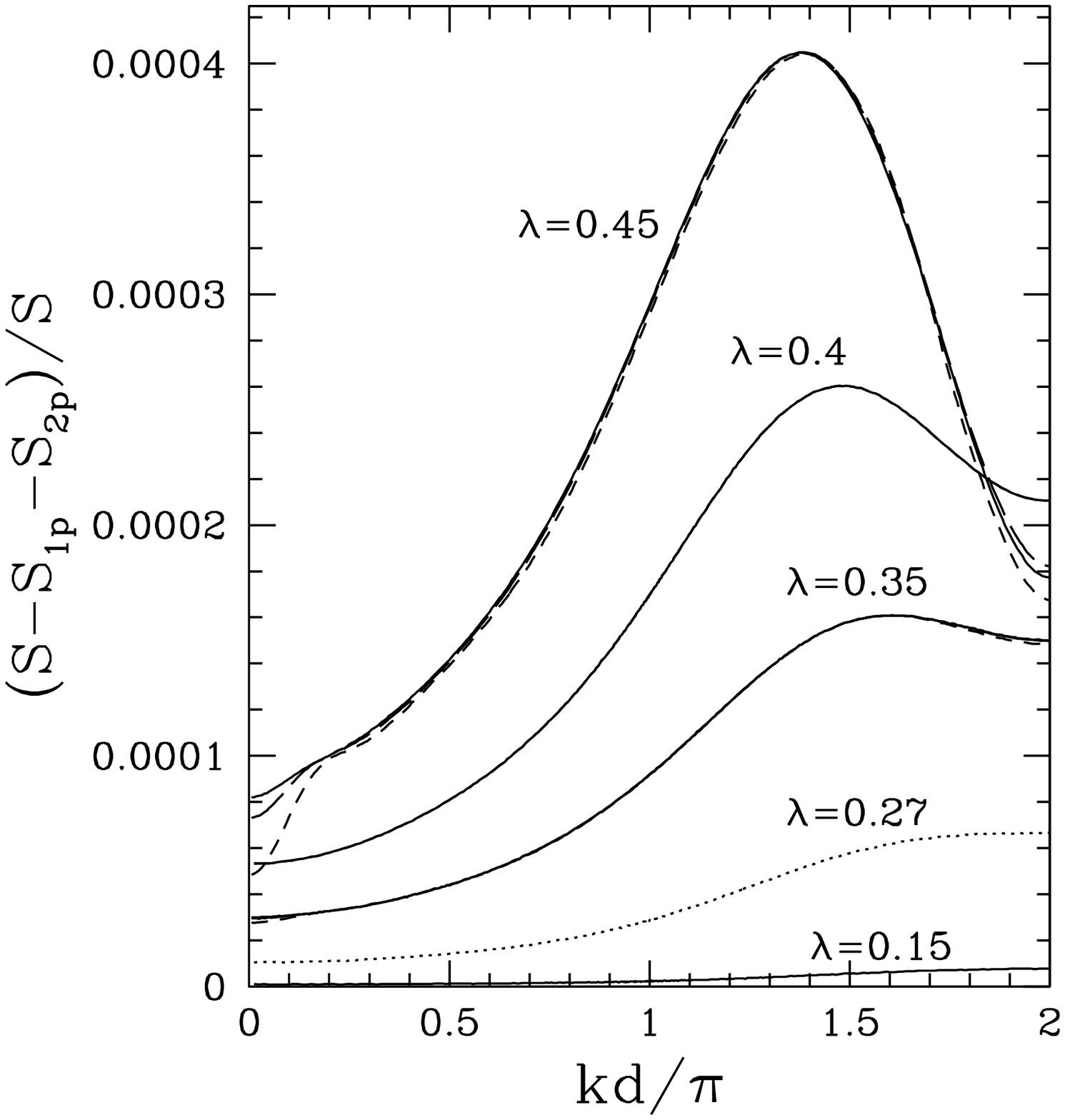,width=9cm}
  \vskip 5mm
 \caption[]
         {The remaining spectral weight for states of more than 2-particles
         ($(S-S_{\rm 1p}-S_{\rm 2p})/S$)  for
         $\lambda=0.15$, 0.27, 0.35,  0.4, 0.45.
         The results of the three highest orders are plotted.}
 \label{fig_Smp_o_S_k}
 \end{center}
\end{figure}

Subtracting the weight of 1 and 2-particle states from
the total weight,
one can obtain the remaining weight $(S-S_{\rm 1p} - S_{\rm 2p})/S$
for states of more than two-particles.
The results for various $\lambda$ are shown in
Fig. \ref{fig_Smp_o_S_k}. For small $\lambda$, the remaining weight
increases as $k$ increases, while for larger $\lambda$, it develops a peak near
$kd=1.5 \pi$, and the remaining weight near $kd=2\pi$ decreases as
$\lambda$ increases.
These weights are tiny, as can be seen, but increasing with $\lambda$.
% These results agree with those of Schmidt and Uhrig\cite{sch02}, who found
% that the 2-particle states dominate the spectral weight, even in the limit $\lambda\to 1$.

\subsubsection{The individual weights for triplet bound states and continuum}

For the bound state $T_1$, we can obtain an
analytic expression for its structure factor
up to order $\lambda^3$
\bea
S_{T_1} (kd) &=&
- {\lambda}^2\,\left( 1 + 2\,\cos (kd) \right) \,{ \sin^6 (\frac{kd}{4})}
- \frac{{\lambda}^3}{24}
{\Big (} -60 + 101\,\cos (\frac{kd}{2}) - 102\,\cos (kd) \nonumber \\
&& + 69\,\cos (\frac{3\,kd}{2}) -
   48\,\cos (2\,kd) + 22\,\cos (\frac{5\,kd}{2}) {\Big )} \,{\sin^4 (\frac{kd}{4})}
   + O(\lambda^4) \label{eqST1}
\eea
where the $\lambda^2$ term agrees with that obtained by Barnes {\it et al}.\cite{ten02}

As a byproduct of our calculations, we also get
an analytic expression for the coherence length\cite{zhe01} which is defined as
\be
L = {\sum_{{\bf r}} \vert {\bf r} \vert f_{{\bf r}}^2\over \sum_{{\bf r}} f_{{\bf r}}^2}
\ee
where $f_{{\bf r}}$ is the amplitude (the eigenvector) for two single-particle excitations
 separated by distance ${\bf r}$ (see Eq. \ref{eq27}). The result up to order $\lambda^2$ for bound state $T_1$
is
\bea
\frac{1}{L_{T_1}} &=&
-1 - 2\,\cos (kd) + \frac{\lambda}{2} \,\left[ -24 - 36\,\cos (kd) - 17\,\cos (2\,kd)
 - 5\,\cos (3\,kd) \right]
 + \frac{{\lambda}^2}{32}
{\Big [} -3410 \nonumber \\
&& - 5937\,\cos (kd) - 3768\,\cos (2\,kd) - 1619\,\cos (3\,kd)
- 482\,\cos (4\,kd) - 100\,\cos (5\,kd) {\Big ]}
+ O(\lambda^3) \label{eq_LT1}
\eea

With these expressions, one can determine the critical momentum $k_c$
where  $S_{T_1}$ or $1/L_{T_1}$ vanishes.
For both $S_{T_1}$ or $1/L_{T_1}$, one gets the same $k_c$ as
\be
k_cd = \left\{ \begin{array}{ll}
2 \pi/3 + 5 \lambda/(4 \sqrt{3}) - 757 \lambda^2/(192 \sqrt{3}) + O(\lambda^3),\quad & k_cd < \pi  \\
4 \pi/3 - 5 \lambda/(4 \sqrt{3}) + 757 \lambda^2/(192 \sqrt{3}) + O(\lambda^3), \quad & k_cd>\pi
\end{array}
\right.
\label{eqKc}
\ee
This agrees with previous results obtained from the
two-particle binding energy\cite{zhe01a}. The expressions (\ref{eqST1})
and (\ref{eq_LT1}) are valid within these regions of momentum.
In the limit $k\to k_c$, the behaviours of $S_{T_1} (k)$ or $1/L_{T_1}$  are
\be
S_{T_1} (k) = \left\{ \begin{array}{ll}
   (k-k_c)d \,{{\lambda}^2}\,\left( 24 + 29\,\lambda \right) /
    (512\,{\sqrt{3} } )
    + O(\lambda^4),\quad & k_cd < \pi  \\
 9\,{\sqrt{3}}\, (k_c-k)d \,\left( 24 - 31\,\lambda \right) \,
      {{\lambda}^2} /512
      + O(\lambda^4), \quad & k_cd>\pi
\end{array}
\right.
\label{eqSKc}
\ee

\be
1/L_{T_1} =
\frac{ |k-k_c|d \,\left( 144 + 12\,\lambda - 443\,{\lambda}^2 \right) }{48\,{\sqrt{3}}}
\ee

So as $k\to k_c$, $S_{T_1} (k)$ and $1/L_{T_1}$ are proportional to $(k-k_c)$,
whereas the binding energy is proportional to $(k-k_c)^2$\cite{zhe01a}.

Integrating  $S_{T_1}$ in Eq. (\ref{eqST1})
over the momenta given in Eq. (\ref{eqKc}), one can get the
auto-correlation function
for $T_1$ as
\be
\Phi_{T_1} =
{\lambda}^2\,\left( -\left( \frac{1}{6} \right)  + \frac{23\,{\sqrt{3}}}{64\,\pi } \right)
+ {\lambda}^3\,\left( \frac{433}{576} - \frac{361\,{\sqrt{3}}}{256\,\pi } \right)
  + O(\lambda^4)
\ee

Since the bound state $T_2$ only appears at $kd =\pi$ in the small $\lambda$ limit,
we cannot get a similar analytic expression for it,
nor for the 2-particle continuum. But the series for exclusive 2-particle  matrix elements
$ \Omega_{\Lambda}^{\rm 2p} $  have been computed
up to order  $\lambda^{12}$, and with this one can compute numerical
results, for any given value of $k$, for the exclusive structure
factor of the 2-particle triplet bound states ($T_1$ and $T_2$)
and 2-particle continuum through  the
finite lattice approach\cite{zhe01a}. We can also get
some series in $\lambda$ for the structure
factor  over those momenta
$k$ where the bound states appear in the limit $\lambda\to 0$,
and the series for $kd=\pi$ are given in Table \ref{tab4}.

\begin{figure}
\begin{center}
% \vskip -7mm
 \epsfig{file=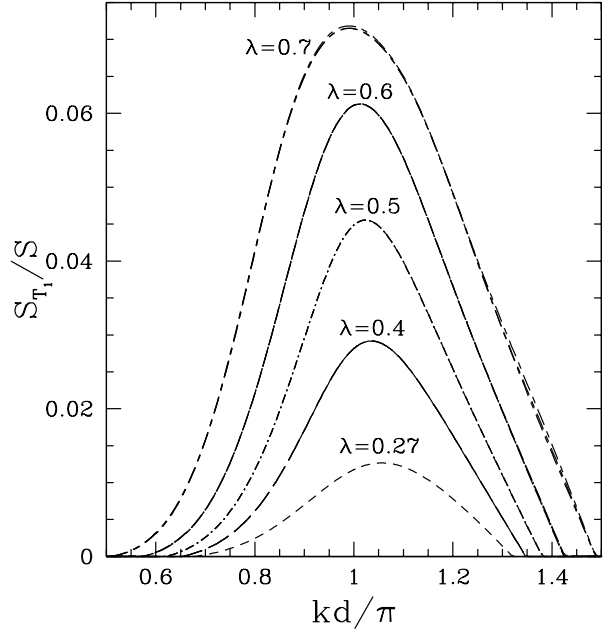,width=9cm}
  \vskip 5mm
 \caption[]
         {The relative weight for bound state $T_1$ ($S_{T_1}/S$)  for
         $\lambda=0.27$,  0.4, 0.5,  0.6, 0.7.
         The results of the three highest orders are plotted.}
 \label{fig_St1_o_S_k}
 \end{center}
\end{figure}

\begin{figure}
\begin{center}
% \vskip -7mm
 \epsfig{file=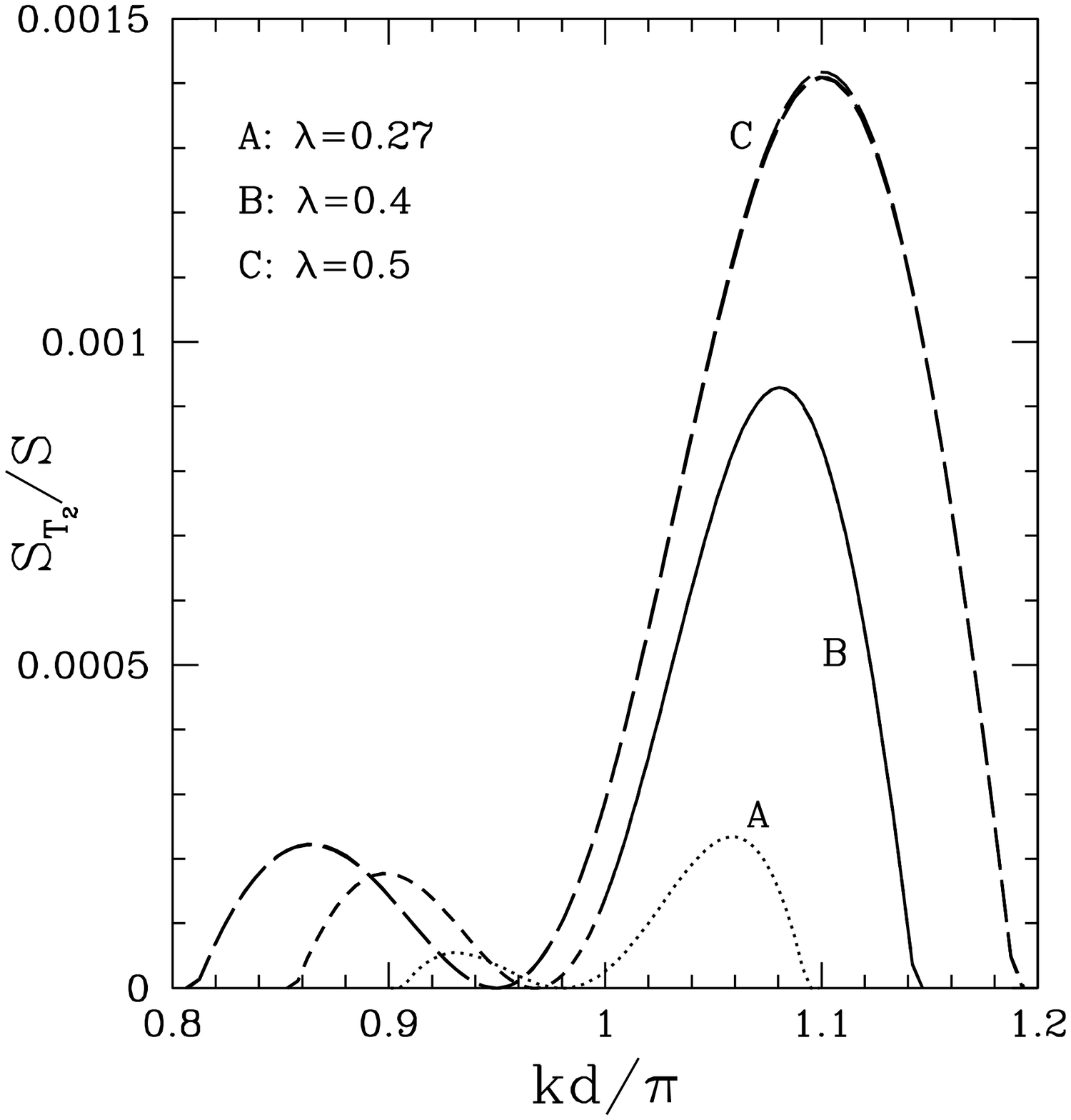,width=9cm}
  \vskip 5mm
 \caption[]
         {The relative weight for bound state $T_2$ ($S_{T_2}/S$)  for
         $\lambda=0.27$,  0.4, 0.5.
         The results of the three highest orders are plotted.}
 \label{fig_St2_o_S_k}
 \end{center}
\end{figure}

\begin{figure}
\begin{center}
% \vskip -7mm
 \epsfig{file=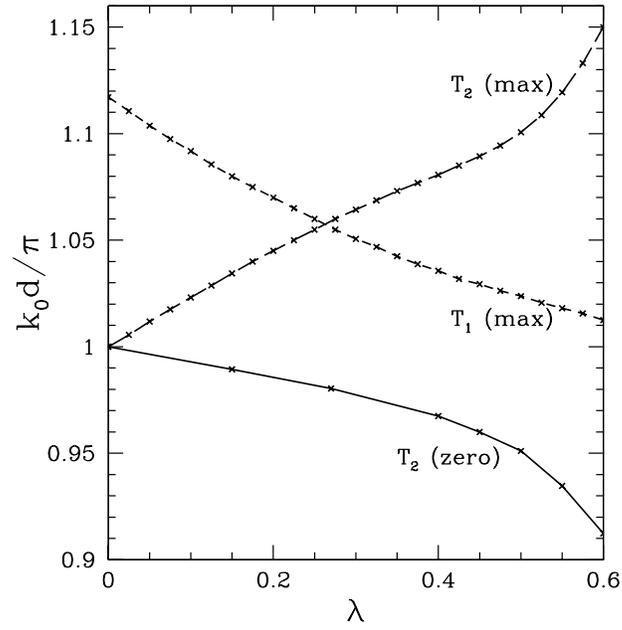,width=9cm}
  \vskip 5mm
 \caption[]
         {The momentum $k_0d/\pi$ or $k_cd/\pi$ where $T_1$ ($T_2$) has its maximum
         (respectively zero and maximum) relative
         spectral weight graphed versus $\lambda$.}
 \label{fig_Kc_T2}
 \end{center}
\end{figure}

The relative spectral weights for $T_1$ and $T_2$ versus $k$ are shown in
Figs. \ref{fig_St1_o_S_k} and \ref{fig_St2_o_S_k}
for several values of $\lambda$.
The weight for $T_1$ is nonzero only over a finite range of momentum where the
bound state exists. In the limit $\lambda\to 0$,
$S_{T_1}/S$ has a maximun at $kd=
4\,\arccos {\big (} {\sqrt{ (5- \sqrt{3} )/8 }} {\big )}=
1.117 \pi$, and as $\lambda$ increases,
the maximum position $k_0d$ moves towards to $k_0d=\pi$. The results for $k_0d$
as function of $\lambda$ are given
in Fig. \ref{fig_Kc_T2}.
The relative spectral weight for $T_2$ exhibits an interference zero near $kd\sim \pi$,
which can be traced back to the
bound-state wave function for $\lambda \to 0$, where the two triplets
are separated by an odd number of dimers\cite{zhe01a}.
The position of this zero shifts to smaller $k$ as $\lambda$ increases, and
is also shown in Fig. \ref{fig_Kc_T2}.
% two peaks, with the peak at larger
% $k$  substantially larger than the one at smaller $k$.
% Between these two peaks, there is a special point at $k_0$ with zero spectral
% weight. From the graph, we can see that $k_0$ shifts to smaller values as
% $\lambda$ increases. The results for $k_0$ as function of $\lambda$ are given
% in Fig. \ref{fig_Kc_T2} (also shown in this figure is the location of
% the maximum relative spectral weight for $T_2$).
 Another interesting feature here is that
 the spectral weight vanishes like $(k-k_c)^2$ as $k\to k_c$, rather than
$(k-k_c)$ as we have seen in Eq. (\ref{eqSKc}).
Also shown in Fig. \ref{fig_Kc_T2} is the location of the maximum relative
weight for $T_2$.

% The existence of a double peak is somehow related to the wavefunction
% of $T_2$. From the wavefunction for small $\lambda$ given in
% Ref. \cite{zhe01a}, we can see that the series for
% $S_2$ appears to be nonzero
% from order $\lambda^4$ for $kd=\pi$, while it become nonzero
% from order $\lambda^2$ for $kd\neq \pi$; but physical reason
% why $T_2$ has zero spectral weight between the two peaks with
% nonzero binding energy is still somehow unclear(?, keep it to
% see any comment from Rajiv).

\begin{figure}
\begin{center}
% \vskip -7mm
 \epsfig{file=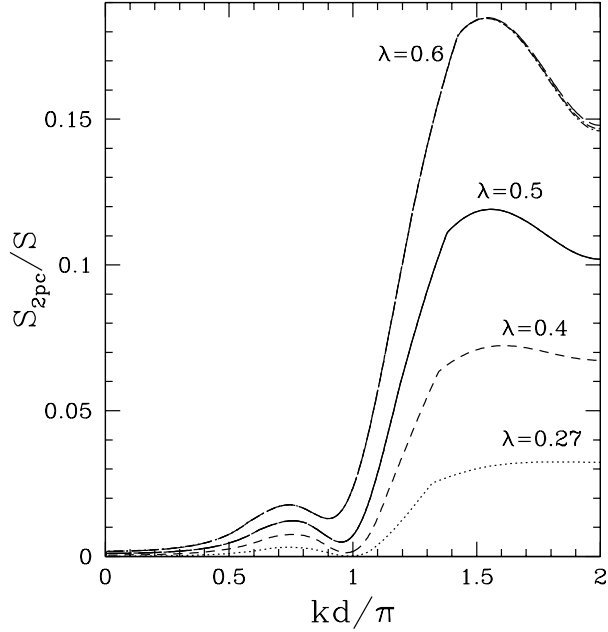,width=9cm}
  \vskip 5mm
 \caption[]
         {The relative weight for the 2-particle continuum ($S_{\rm 2pc}/S$)  for
         $\lambda=0.27$,  0.4, 0.5, 0.6.
         The results of the three highest orders are plotted.}
 \label{fig_S2pc_o_S_k}
 \end{center}
\end{figure}

The remaining weight for the 2-particle continuum $S_{\rm 2pc}$ is
shown in Fig. \ref{fig_S2pc_o_S_k}. These curves show a dip near $kd=\pi$,
because the bound states absorb some of the weight in that region.
%We can also compute $S_{\rm 2pc}$ as function of $k$ and $\omega$:
%the results for $\lambda =0.27$ were presented in our short paper\cite{zhe02},
%and we will not present any further results here.

%\subsection{Center and Width}
\subsection{Complete Dynamical Structure Factor}

\begin{figure}
 \begin{center}
 \vskip -7mm
 \epsfig{file=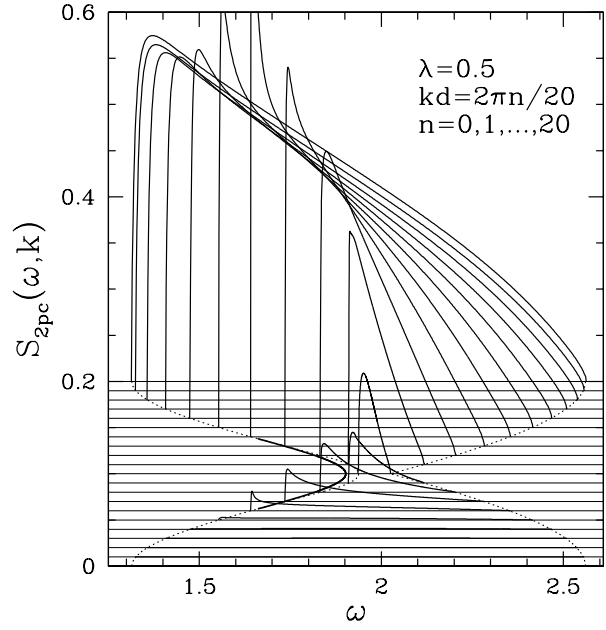, width=9cm}
  \vskip 5mm
 \caption[]
         {The structure factor (shifted by $n/100$) for the 2-particle continuum versus energy $\omega$
          at $\lambda=0.5$ and $kd=2\pi n/20$, $n=0,1,2,\cdots,20$, Also shown as a bold
          solid line is the dispersion relation for the triplet bound-state $T_1$.}
 \label{fig_weight_2pc_p5}
 \end{center}
\end{figure}

The complete dynamical structure factor for $\lambda=0.27$
was presented in our earlier paper\cite{zhe02}. The dynamical structure
factor for the 2-particle continuum at $\lambda=0.5$ is given in Fig. \ref{fig_weight_2pc_p5}.
A notable feature is the spike which develops as a bound state enters the continuum, discussed more
fully in Ref.  \onlinecite{zhe02}.
We will be happy to provide the complete
dynamical structure factor for the AHC model for any value of
$\lambda$, if requested.

\begin{figure}
\begin{center}
% \vskip -7mm
 \epsfig{file=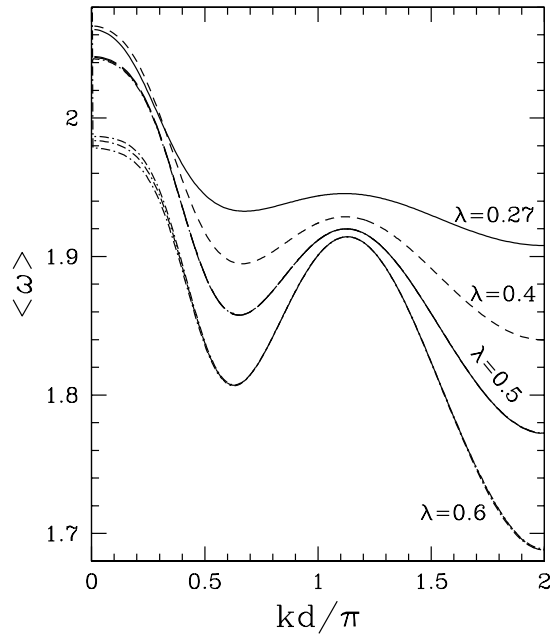,width=9cm}
  \vskip 5mm
 \caption[]
         {The 2-particle energy centroid
$\langle \omega \rangle$  versus $k$
          for $\lambda=0.27$,
          0.4, 0.5, 0.6.}
 \label{fig_av_E}
 \end{center}
\end{figure}

\begin{figure}
\begin{center}
% \vskip -7mm
 \epsfig{file=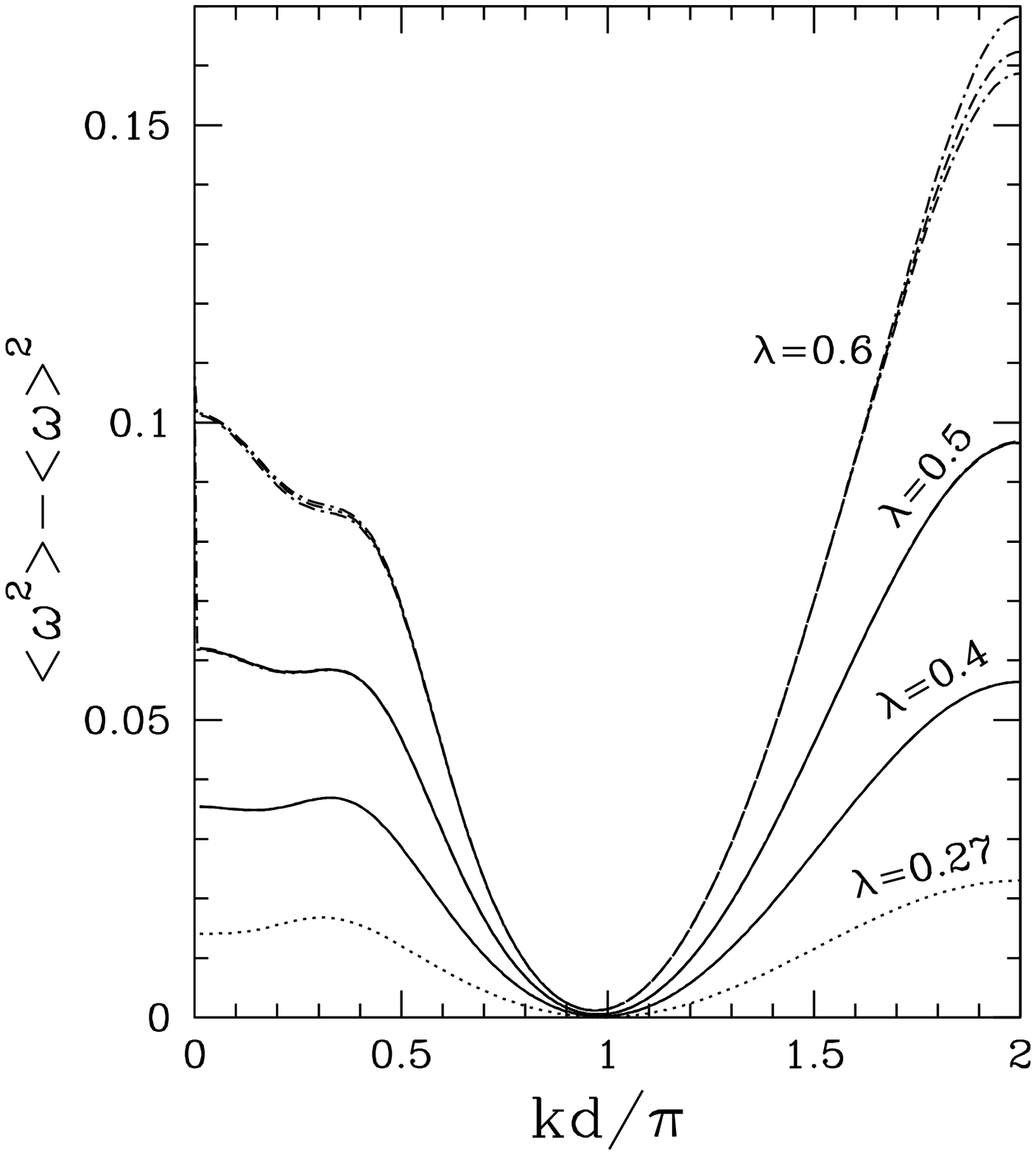,width=9cm}
  \vskip 5mm
 \caption[]
         {The 2-particle
width $\langle \omega^2 \rangle - \langle \omega \rangle^2$  versus $k$
          for $\lambda=0.27$,
          0.4, 0.5, 0.6.}
 \label{fig_width}
 \end{center}
\end{figure}

Given the structure factor, one can also compute, for 2-particle states, the
energy centroid
$\langle \omega \rangle$ and the
width $\Delta \omega$ which are  defined as
\bea
\langle \omega \rangle &=& { \int S_{2} (\omega, k) \omega d\omega \over
\int S_{2} (\omega, k) d\omega } \\
\Delta \omega &=& \langle \omega^2 \rangle - \langle \omega \rangle^2
\eea
where $S_2 (\omega, k)$ is the structure factor of the 2-particle states, including
both the 2-particle bound states and 2-particle continuum.
% (Note: I am not sure whether we should use the structure factor of
% 2-particle continuum only here,
% the results will be difference from the above definition on the region
% near $k=\pi$ where the bound states exist)
These quantities can be extracted readily from experiments. Our results
are given in Figures \ref{fig_av_E} and \ref{fig_width} for various $\lambda$.
 We can see that $\Delta \omega$ has a nonzero minimum near $kd=\pi$, reflecting the presence
of  strong 2-particle bound states and a weak and narrow 2-particle continuum.

\subsection{Auto-correlation function}
Finally, we discuss the results for the spin auto-correlation functions.
First, let us discuss the critical behaviour as $\lambda \to 1$.
Schmidt and Uhrig\cite{sch02} argued that the critical behaviour for
the total auto correlation function (summed over $\omega$) of
the 1-particle state $\Phi_{\rm 1p}$ and 2-particle states $\Phi_{\rm 2p}$ should be
\bea
\Phi_{\rm 1p} &\propto & (1-\lambda)^{1/3}  \label{C_Phi1p}  \\
\Phi_{\rm 2p} &\propto & {\rm const.} + O( (1-\lambda)^{1/3} )
   \label{C_Phi2p}
\eea
modulo logarithms, as for the structure factors.

To verify these behaviours, we have applied Dlog Pad\'e
approximants to the series, and the results obtained from
unbiased and biased Dlog Pad\'e approximants are shown in Table \ref{tab5}.
For  $\Phi_{\rm 1p}$, the results are consistent
with Eq. (\ref{C_Phi1p}).
Assuming this behaviour,  we can estimate the prefactor
using integrated differential approximants. The result is
\be
\Phi_{\rm 1p} = 1.258(2)(1-\lambda)^{1/3}
\ee

For $\Phi_{\rm 2p}$, however,
the results are more ambiguous.
The unbiased Dlog Pad\'e approximants to $\partial \Phi_{\rm 2p}/\partial\lambda$
 tend to exhibit defects
for $\lambda<1$, and the biased critical index for
$\partial \Phi_{\rm 2p}/\partial\lambda$ is about -0.6,
slightly smaller than -2/3. This might be due to an extra logarithmic correction.

\begin{figure}
\begin{center}
% \vskip -7mm
 \epsfig{file=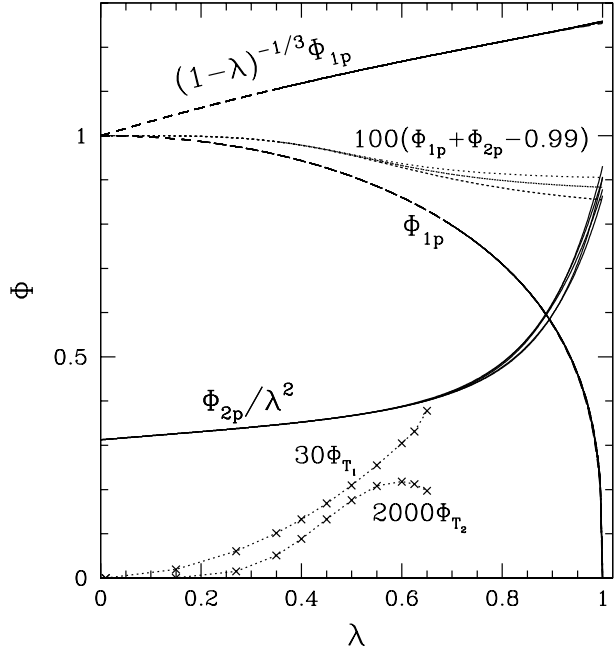,width=9cm}
  \vskip 5mm
 \caption[]
         {The auto correlation functions versus $\lambda$
         for the 1-particle state ($\Phi_{\rm 1p}$),
         2-particle states  ($\Phi_{\rm 2p}$), and two particle bound states
         $T_1$ and $T_2$.}
 \label{fig_autoC}
 \end{center}
\end{figure}

The various auto-correlation functions versus $\lambda$ are shown in Fig. \ref{fig_autoC},
where one can see that $\Phi_{\rm 1p}$ vanishes at the limit
$\lambda=1$, while $(1-\lambda)^{-1/3} \Phi_{\rm 1p}$ increases almost linearly
as $\lambda$ increases.
The curve for $(\Phi_{\rm 1p} +\Phi_{\rm 2p}) $,
if we {\it assume} it is non-singular at $\lambda=1$ (i.e. the
singularities in $\Phi_{\rm 1p}$ and $\Phi_{\rm 2p}$ cancel exactly),
runs almost flat with $\lambda$ once we neglect unphysical and defective
approximants: that would indicate that the 2-particle sector accounts for
about 99.8\% of the weight, even at $\lambda=1$,
which agrees almost exactly with the conclusions of Schmidt and
Uhrig\cite{sch02}. Remarkably, this is much higher than the
fraction of 73\% for the
two-spinon continuum at $\lambda=1$ calculated by Karbach {\it et al.}\cite{kar97}
from the exact solution.
The result that sectors with more than 2 particles account for very little weight
can be further understood from  Fig. \ref{fig_Smp_o_S_k}, where we can see that
the remaining spectral weight near $kd=2\pi$
 for states of more than 2-particles
actually decreases with $\lambda$, for large $\lambda$.
Also shown in Fig. \ref{fig_autoC} is the direct
extrapolation of the 2-particle auto-correlation $\Phi_{\rm 2p}$ using
integrated differential approximants. These extrapolations assume that
there is {\it no}
singularity in $\Phi_{\rm 2p}$ at $\lambda=1$, and the results
give a somewhat smaller value of about 0.9  at $\lambda=1$.
%
% The 2-particle auto correlation function
%  $\Phi_{\rm 2p}$ is about 0.86(10) at $\lambda=1$: this is
% larger than the 73\% estimated by Karbach {\it et al.}\cite{kar97},
% but smaller than the recent estimate of 99\%
% by Schmidt {\it et al.}\cite{sch02} with shorter series.
% Schmidt {\it et al.}
% probably overestimate this because they have assumed the
% critical form Eq. (\ref{C_Phi2p}). If we used this critical
% form, we would get $\Phi_{\rm 2p}$ larger than 1.
%
For the bound states, the auto  correlation function for $T_2$ increases, and reaches its maximum
around $\lambda=0.6$, then decreases, while the auto  correlation function for $T_1$
continues to increase as far as we can follow it. We presume that $\Phi_{T_1}$
will also vanish as $\lambda\to 1$.

\begin{figure}
\begin{center}
% \vskip -7mm
 \epsfig{file=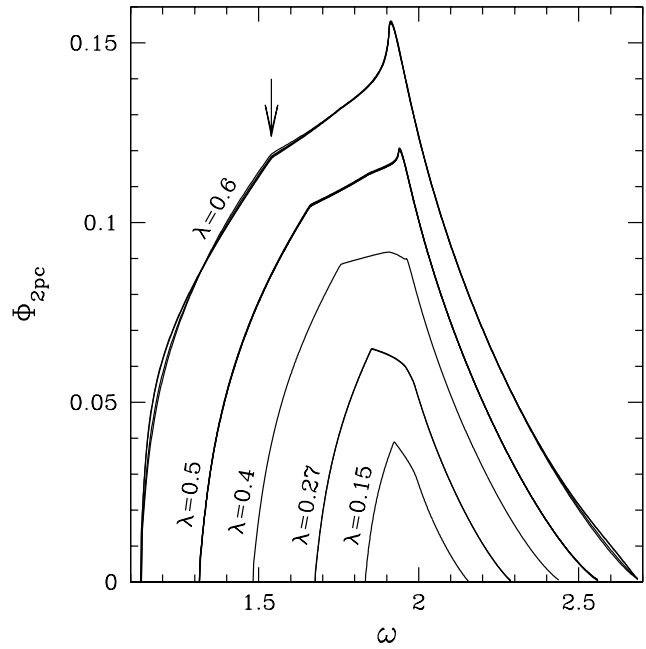,width=9cm}
  \vskip 5mm
 \caption[]
         {The auto correlation function versus energy $\omega$
         for the  2-particle continuum ($\Phi_{\rm 2pc}(\omega)$) for $\lambda=0.15$,
         0.27, 0.4, 0.5, 0.6.}
 \label{fig_autoC_2pc}
 \end{center}
\end{figure}

The autocorrelation function
for the 2-particle continuum ($\Phi_{\rm 2pc}(\omega)$) is shown
for various $\lambda$ in Figure \ref{fig_autoC_2pc}.
The major feature is a spike at $\omega \sim 2$ which
becomes more prominent as $\lambda$ increases. This
spike is due to contributions from the region around $kd=\pi$ near
the lower threshold of the dispersion curve. It becomes
divergent as $\lambda\to 1$, and matches rather neatly with
a logarithmic divergence
at $\omega=\pi/2$ discovered in the 2-spinon continuum
by Karbach {\it et al.}\cite{kar97}.
Unfortunately, we can not compute an explicit series
for $\Phi_{\rm 2pc}(\omega)$, so no series extrapolation can be made here.

Figure \ref{fig_autoC_2pc} also displays another small cusp or spike at lower
$\omega$, which occurs at the threshold energy where the triplet bound state $T_1$ merges
with the continuum, i.e. where the structure factor diverges as shown in Figure
\ref{fig_weight_2pc_p5}. This position is marked by the arrow in Figure
\ref{fig_autoC_2pc}, for the case $\lambda=0.6$. As $\lambda\to 1$, the $T_1$ threshold
migrates towards $\omega =0$, and the logarithmic divergence there\cite{kar97} at
$\lambda=1$ may be viewed as relic of the $T_1$ bound state.

\begin{figure}
\begin{center}
% \vskip -7mm
 \epsfig{file=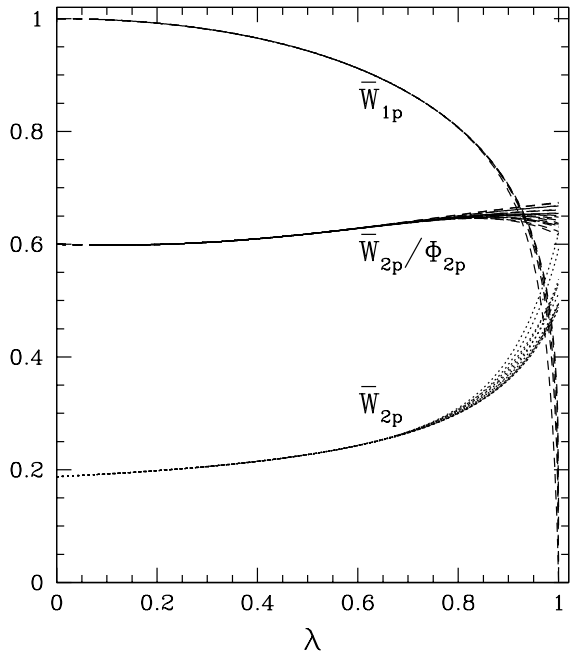,width=9cm}
  \vskip 5mm
 \caption[]
         {The average weights ($\overline{W}_{\rm 1p}$ and $\overline{W}_{\rm 2p}$)
          for 1 and 2 particle states, and ratio
         $\overline{W}_{\rm 2p}/\Phi_{\rm 2p}$ versus $\lambda$.
         Several different integrated differential approximants to
the series are shown.}
 \label{fig_av_W}
 \end{center}
\end{figure}

Note that the structure factor is dominated for $\lambda\to 1$
by $kd=2\pi$, and so
is the auto-correlation function. To get an idea of the multiparticle
contributions to the dynamical structure factor at all wavevectors, we
define the {\it average} relative weight, for a particular state $\Lambda$, as
\be
\overline{W}_{\Lambda} = {1\over 2 \pi} \sum_{\omega} \int_0^{2\pi}
{S_{\Lambda} (k,\omega) \over S(k)} dk
\ee
The series for the average relative weight for  2-particle states  is
given in Table \ref{tab4}, and graphed as function of $\lambda$ in
Figure \ref{fig_av_W}. Once again, we see that the average weight for the
1-particle state drops to zero as $\lambda \to 1$. The average weight
for the 2-particle states has a sharp increase near $\lambda=1$ which
makes it difficult to estimate the results at $\lambda=1$, but the result
for the ratio $\overline{W}_{\rm 2p}/\Phi_{\rm 2p}$ is quite flat.
Hence one estimates that $\overline{W}_{\rm 2p}$
remains substantial, at about 60\%,
as $\lambda \to 1$. This is smaller than $\Phi_{\rm 2p}$, showing
that multiparticle excitations are more important away from the
antiferromagnetic wavevector $kd=2\pi$.

A priori, one might have expected that the average weights for 1-particle states, 2-particle states,
etc, would all tend to zero as $\lambda\to 1$, with all the weight moving
into many-particle states. Instead of that, we find that the weight for the
2-particle continuum remains finite and large, comparable to or
by some measures
even greater than that of the 2-spinon continuum computed by
Karbach {\it et al.}\cite{kar97}

\section{Conclusions and Discussion}

We have shown in this paper how to calculate multiparticle structure
factors and spectral weights
to high orders in perturbation theory using linked-cluster
expansion techniques\cite{tre00,zhe01}. Applying these techniques
to the case of the alternating Heisenberg chain, a detailed picture
has been given of both the integrated structure factors, and the individual
spectral weights for the 1-particle state, 2-particle bound states and 2-particle
continuum as functions of wavevector $k$.
Continuing the series by means of Pad\'e approximants or integrated
differential approximants, good convergence is obtained from $\lambda=0$ right
up to $\lambda=1$.
Hopefully, it should be possible to test these predictions against experiments
in the near future\cite{xu00,ten02}.

%An alternative technique is the `continous unitary transformations' (CUTS)
%method\cite{kne01} used by Schmidt and  Uhrig\cite{sch02} to calculate
%integrated structure factors for the model. Our results are in general
%agreement with theirs, but are  carried to higher orders and for many more
%quantities.

The 1-particle energy gap and spectral weight at general momenta appear to
vanish as $\lambda\to 1$, following the behaviour predicted by Cross and Fisher\cite{den79},
and already confirmed numerically by Singh and Zheng\cite{sin99}. This
would seem to confirm the general notion  that the triplets no longer form elementary
excitations for the system at $\lambda=1$.
However, the 2-triplet spectral weight remains finite in the uniform limit
and, in fact, appears to form the major part of the
total spectral weight\cite{sch02}. Schmidt and
Uhrig\cite{sch02} already pointed out that indeed the 2-triplet states carry a larger portion
of the total spectral weight than the 2-spinon states, calculated by Karbach
{\it et al.}\cite{kar97}. We also find that the 2-particle
auto-correlation functions for triplets display singularities similar
to the known singularities for spinons\cite{kar97}.

We find that clear precursors to many features of the two-spinon
continuum for the Heisenberg model, such as vanishing weight at the
upper end of the continuum, divergent weight at the lower end of the
continuum, cusps in the autocorrelation function, {\it etc.}, are
already evident in the two-triplet continuum of the alternating chain.
We have discovered an interesting result in that the logarithmic divergences
at $\omega=0$
in the auto correlation function of the uniform chain appears to be a relic
of the triplet bound state $T_1$ in the non-uniform chain.
Our  highly accurate calculations of frequency and
wavevector resolved spectral-weights should prove useful
in understanding the spectral functions of real materials.

The crossover from elementary triplets to spin-half elementary excitations
is quite different here than in the case of the $J_1-J_2-\delta$ model with $J_2/J_1>0.24$, where
the uniform limit stays spontaneously dimerized.
Affleck {\it et al}\cite{sor98} presented the following picture in the latter
case. The mass gap decreases exponentially as the uniform limit
is approached, and in that limit the low-lying spectrum consists of a gapless continuum of
soliton-antisoliton states. Away from the uniform limit, the soliton-antisoliton states are
confined by a linear potential, giving rise to a ladder of discrete states, which
correspond to the triplet excitations and their bound states. In the triplet languange,
on the other hand, the triplet energy gap drops to zero as the uniform limit is
approached, and the multi-triplet bound states condense to form a continuum matching the
soliton-antisoliton description. Numerical evidence confirms this picture\cite{sor98,zhe01a}.

The scenario appears to be quite different in the present case, where there is no next-nearest-neighbour
interaction ($J_2$). Here the dimerization interaction has dimension $\frac{1}{2}$,
giving rise to a gap which vanishes as $\delta^{2/3}$ in the uniform limit.
In the soliton language\cite{das75}, at small but non-zero $\delta$, the low-lying spectrum
consists of a soliton, an antisoliton, and a soliton-antisoliton bound state,
forming a degenerate triplet, plus just one other soliton-antisoliton state at
$\sqrt{3}$ times the mass. Since the soliton and antisoliton are not confined,
a soliton-antisoliton continuum occurs at higher energy. In the triplet language,
the 1-particle triplet corresponds to the three degenerate states, and the singlet
bound state corresponds to the higher-lying one.
There is no
condensation of multi-triplet bound states; instead, the 2-triplet continuum drops
down to match the spinon-antispinon continuum. The mass and spectral weight of the
1-triplet state vanishes as the uniform limit is approached; and naively one might
expect the same to happen for the 2-triplet, 3-triplet states, etc, leaving many-triplet
states to correspond to the spinon-antispinon continuum. Instead, we have found that the
2-triplet states dominate the total weight in the uniform limit, in agreement with
Schmidt and Uhrig\cite{sch02}.
This is a peculiar and paradoxical feature, which argues that the
triplet description remains at least as relevant as the spinon description,
even in the uniform limit. The multi-triplet
states ($n>2$) still appear to carry only a tiny fraction of the spectral
weight at $\lambda=1$.

These findings call for further theoretical interpretation.
Karbach {\it et al.}\cite{kar02} have described the uniform Heisenberg antiferromagnet in
terms of three different elementary excitations - magnons, spinons, and `psinons' -
each of which can be useful in different circumstances. None of these correspond to
our dimer triplet excitations, however. Our work shows that a
conventional picture, built as a perturbation around a trivial limit,
can provide a highly accurate quantitative
description of the system, when carried out to high orders.
%A bosonized effective Hamiltonian for the
%triplets has been discussed\cite{sor98,jia01}, but the behaviour of the multi-triplet states
%has not been analyzed.

\begin{acknowledgments}
This work is supported in part by a grant
from the Australian Research Council and the US National Science Foundation
grant number DMR-0240918. We have benefited from discussions with
Professor O.P. Sushkov and Professor J. Oitmaa.
The computation has been performed on the AlphaServer SC
 computer. We are grateful for the computing resources provided
 by the Australian Partnership for Advanced Computing (APAC)
National Facility.
\end{acknowledgments}

%\appendix
%\section{Appendixes}

%\newpage
\bibliography{basename of .bib file}

\begin{references}
\bibitem[*]{cjh} Email address: c.hamer@unsw.edu.au
\bibitem[\dag]{zwh} Email address: w.zheng@unsw.edu.au
\bibitem{tre00} S. Trebst, H. Monien, C.J. Hamer, W. Zheng, R.R.P.
Singh, Phys. Rev. Lett. {\bf 85}, 4373 (2000).

\bibitem{zhe01} W. Zheng, C.J. Hamer, R.R.P. Singh, S. Trebst, H.
Monien, Phys. Rev. {\bf B63}, 144410 (2001).

\bibitem{kne01} C. Knetter, K.P. Schmidt, M. Gr{\" u}ninger and G.S.
Uhrig, Phys. Rev. Lett. {\bf 87}, 167204 (2001).

\bibitem{gel00}M.P. Gelfand and R.R.P. Singh, Adv. Phys. {\bf 49}, 93(2000).

\bibitem{xu00}
G. Xu, C. Broholm, D.H. Reich and M.A. Adams, Phys. Rev. Lett. {\bf 84},
4465 (2000).

\bibitem{ten02} D.A. Tennant, C. Broholm, D.H. Reich, S.E. Nagler, G.E.
Granroth, T. Barnes, K. Damle, G. Xu, Y. Chen and B.C. Sales,
cond-mat/0207678

\bibitem{uhr96} G.S. Uhrig and H.J. Schulz, Phys. Rev. {\bf B54}, R9624
(1996).

\bibitem{bou98}
G. Bouzerar, A.P. Kampf and G.I. Japaridze, Phys. Rev. {\b58}, 3117
(1998).

\bibitem{fle97}
A. Fledderjohann and C. Gros, Eur. Phys. Lett. {\bf 37}, 189 (1997).

\bibitem{zhe01a} W. Zheng, C.J. Hamer, R.R.P. Singh, S. Trebst, H.
Monien, Phys. Rev. {\bf B63}, 144411 (2001).

% 2. Critical behaviour
\bibitem{den79}
M.P.M. den Nijs, Physica {\bf 95A}, 449 (1979).
M.C. Cross and D. Fisher, Phys. Rev. {\bf B19}, 402 (1979).
J.L. Black and V.J. Emery, Phys. Rev. {\bf B23}, 429 (1981).

\bibitem{uhr99}
G.S. Uhrig, F. Sch{\"o}nfeld, M. Laukamp and E. Dagotto, Eur. Phys. J.
{\bf B7}, 67 (1999).
T. Papenbrock, T. Barnes, D.J. Dean, M.V. Stoitsev and M.R. Stayer, cond-mat/0212254.


% 3. Triplet/spinon crossover:
\bibitem{sor98}
E.S. Sorenson, I. Affleck, D. Augier and D. Poilblanc, Phys. Rev. {\bf
B58}, R14701 (1998).
%
%\bibitem{aff97}
I. Affleck, in {\it Dynamical Properties of Unconventional Magnetic
Systems} (NATO ASI, Geilo, Norway, 1997).


\bibitem{mar71}
W. Marshall and S.W. Lovesey, {\it Thermal Neutron Scattering}
(Clarendon, Oxford, 1971).


\bibitem{bar99} T. Barnes, J. Riera and D.A. Tennant, Phys. Rev. {\bf
B59}, 11384 (1999).


\bibitem{irv84}
A.C. Irving and C.J. Hamer, Nucl. Phys. {\bf B230}, 361 (1984).


\bibitem{gel90}
M.P. Gelfand, R.R.P. Singh and D.A. Huse, J. Stat. Phys. {\bf 59}, 1093
(1990).

\bibitem{he90}
H-X. He, C.J. Hamer and J. Oitmaa, J. Phys. {\bf A23}, 1775 (1990).

\bibitem{gel96}
M.P. Gelfand, Solid State Commun. {\bf 98}, 11 (1996).

\bibitem{sin95}R.R.P. Singh and M.P. Gelfand, Phys. Rev. B{\bf 52}, R15695(1995).

\bibitem{gel89}
M.P. Gelfand, R.R.P. Singh and D.A. Huse, Phys. Rev. {\bf B40}, 10801
(1989).


\bibitem{sin99} R.R.P. Singh and W. Zheng, Phys. Rev. {\bf B59}, 9911
(1999).


\bibitem{zhe02}W. Zheng, C.J. Hamer and R.R.P. Singh, cond-mat/0211346.


\bibitem{duf68}
W. Duffy and K.P. Bair, Phys. Rev. {\bf 165}, 647 (1968).


\bibitem{bon82}
J. Bonner and H.W.J. Bl{\"o}te, Phys. Rev. {\bf B25}, 6959 (1982).

\bibitem{jia01}
X-F. Jiang, H. Chen and D.Y. Xing, J. Phys. {\bf A34}, L259 (2001).


\bibitem{soo85}
Z.G. Soos, S. Kuwajima and J.E. Mihalick, Phys. Rev. {\bf B32}, 3124
(1985).

\bibitem{spr86}
G. Spronken, B. Fourcade and Y. Lepine, Phys. Rev. {\bf B33}, 1886
(1986).

\bibitem{pap02}T. Papenbrock, T. Barnes, D.J. Dean, M.V. Stoitsov and
M.R. Strayer, cond-mat/0212254.

\bibitem{she99}
P. Shevchenko, V.N. Kotov and O.P. Sushkov, Phys. Rev. {\bf B60}, 31305
(1999).

\bibitem{sch02}K.P. Schmidt, and G.S. Uhrig, Phys. Rev. Lett. {\bf 90}, 227204(2003).

\bibitem{mul02}M. M\"uller and H.J. Mikeska, cond-mat/0211335.

\bibitem{gut}A.J. Guttmann, in ``Phase Transitions and Critical
Phenomena'', Vol. 13 ed. C. Domb and J. Lebowitz (New York, Academic, 1989).


\bibitem{aff98}I. Affleck, J. Phys. A{\bf 31}, 4573(1998).

\bibitem{kar97}M. Karbach, G. M\"uller, and A.H. Bougourzi,
  \prb {\bf 55}, 12510(1997).

\bibitem{das75}R.F. Dashen, B. Hasslacher and A. Neveu, Phys. Rev. D{\bf 11}, 3423(1975).

\bibitem{kar02}M. Karbach, D. Biegel and G. M\"uller, cond-mat/0205142.


% ref. not cited

% 1. Spinon in Heisenberg chain:
% \bibitem{fad81}L.D. Faddeev and L.A. Takhtajan, Phys. Lett. A{\bf 85}, 375(1981).
% F.D.M. Haldane, Phys. Rev. Lett. {\bf 45}, 1358(1980).
% B.S. Shastry and B. Sutherland, Phys. Rev. Lett. {\bf 47}, 964(1981).


%\bibitem{bie02}D. Biegel, M. Karbach and G. M\"uller, Europhys. Lett. {\bf 59}, 882(2002).

%\bibitem{yua00}W. Yuand and S. Haar, Phys. Rev. B{\bf 62}, 344(2000).

%\bibitem{dam98}
%K. Damle and S. Sachdev, Phys. Rev. {\bf B57}, 8307 (1998).

%\bibitem{nis94}
%M. Nishi, O. Fujita and J. Akimitsu, Phys. Rev. {\bf B50}, 6508 (1994).

%\bibitem{ain97}
%M. Ain et al., Phys. Rev. Letts. {\bf 78}, 1540 (1997).

%\bibitem{raj95}R.R.P. Singh and M.P. Gelfand, Phys. Rev. B{\bf 52}, R15695(1995).


\end{references}

% \newpage

\renewcommand{\baselinestretch}{1}

\begin{table}
\squeezetable
\setdec 0.00000000000000
\caption{Series coefficients for the integrated structure factor
$ S(k) = \sum_{n,m} a_{n,m} \lambda^m \cos (n kd/2) $.
Nonzero coefficients $a_{n,m}$
up to order $m=13$ are listed.
  }
\label{tab1}
\begin{tabular}{|rr|rr|rr|rr|}
 \multicolumn{1}{|c}{$(n,m)$} &\multicolumn{1}{c|}{$a_{n,m}$}
& \multicolumn{1}{c}{$(n,m)$} &\multicolumn{1}{c|}{$a_{n,m}$}
& \multicolumn{1}{c}{$(n,m)$} &\multicolumn{1}{c|}{$a_{n,m}$}
& \multicolumn{1}{c}{$(n,m)$} &\multicolumn{1}{c|}{$a_{n,m}$} \\
\tableline
 ( 0, 0) &\dec    1.000000000       &( 4,11) &\dec $-$1.591179666$\times 10^{-3}$ &( 9, 7) &\dec $-$2.661559650$\times 10^{-2}$ &(15, 9) &\dec $-$9.209089311$\times 10^{-3}$ \\
 ( 1, 0) &\dec $-$1.000000000       &( 4,12) &\dec $-$1.346799079$\times 10^{-3}$ &( 9, 8) &\dec $-$1.701711998$\times 10^{-2}$ &(15,10) &\dec $-$1.132484677$\times 10^{-2}$ \\
 ( 1, 1) &\dec $-$2.500000000$\times 10^{-1}$ &( 4,13) &\dec $-$1.135073475$\times 10^{-3}$ &( 9, 9) &\dec $-$1.047345923$\times 10^{-2}$ &(15,11) &\dec $-$1.152471081$\times 10^{-2}$ \\
 ( 1, 2) &\dec    3.125000000$\times 10^{-2}$ &( 5, 2) &\dec $-$9.375000000$\times 10^{-2}$ &( 9,10) &\dec $-$6.243270719$\times 10^{-3}$ &(15,12) &\dec $-$1.055798426$\times 10^{-2}$ \\
 ( 1, 3) &\dec    2.864583333$\times 10^{-2}$ &( 5, 3) &\dec $-$1.119791667$\times 10^{-1}$ &( 9,11) &\dec $-$3.511373171$\times 10^{-3}$ &(15,13) &\dec $-$9.126658607$\times 10^{-3}$ \\
 ( 1, 4) &\dec    1.247829861$\times 10^{-3}$ &( 5, 4) &\dec $-$5.577256944$\times 10^{-2}$ &( 9,12) &\dec $-$1.832316909$\times 10^{-3}$ &(16, 8) &\dec    1.534223557$\times 10^{-3}$ \\
 ( 1, 5) &\dec    4.747178819$\times 10^{-4}$ &( 5, 5) &\dec $-$1.284450955$\times 10^{-2}$ &( 9,13) &\dec $-$8.384814283$\times 10^{-4}$ &(16, 9) &\dec    4.630391871$\times 10^{-3}$ \\
 ( 1, 6) &\dec    2.256346338$\times 10^{-3}$ &( 5, 6) &\dec    1.543539542$\times 10^{-4}$ &(10, 5) &\dec    1.538085938$\times 10^{-2}$ &(16,10) &\dec    7.630811251$\times 10^{-3}$ \\
 ( 1, 7) &\dec    1.115188285$\times 10^{-3}$ &( 5, 7) &\dec    2.104722325$\times 10^{-3}$ &(10, 6) &\dec    2.759165823$\times 10^{-2}$ &(16,11) &\dec    9.315049401$\times 10^{-3}$ \\
 ( 1, 8) &\dec    4.433909083$\times 10^{-4}$ &( 5, 8) &\dec    2.590777271$\times 10^{-3}$ &(10, 7) &\dec    2.762316480$\times 10^{-2}$ &(16,12) &\dec    9.630494889$\times 10^{-3}$ \\
 ( 1, 9) &\dec    4.463708690$\times 10^{-4}$ &( 5, 9) &\dec    2.541851515$\times 10^{-3}$ &(10, 8) &\dec    2.143434446$\times 10^{-2}$ &(16,13) &\dec    9.064708700$\times 10^{-3}$ \\
 ( 1,10) &\dec    3.908585581$\times 10^{-4}$ &( 5,10) &\dec    2.142535125$\times 10^{-3}$ &(10, 9) &\dec    1.509559320$\times 10^{-2}$ &(17, 8) &\dec $-$7.671117783$\times 10^{-4}$ \\
 ( 1,11) &\dec    2.721746108$\times 10^{-4}$ &( 5,11) &\dec    1.784799808$\times 10^{-3}$ &(10,10) &\dec    1.026773209$\times 10^{-2}$ &(17, 9) &\dec $-$2.828093400$\times 10^{-3}$ \\
 ( 1,12) &\dec    2.078935750$\times 10^{-4}$ &( 5,12) &\dec    1.526601703$\times 10^{-3}$ &(10,11) &\dec    6.786844453$\times 10^{-3}$ &(17,10) &\dec $-$5.469609444$\times 10^{-3}$ \\
 ( 1,13) &\dec    1.752731379$\times 10^{-4}$ &( 5,13) &\dec    1.307569342$\times 10^{-3}$ &(10,12) &\dec    4.357777095$\times 10^{-3}$ &(17,11) &\dec $-$7.545021498$\times 10^{-3}$ \\
 ( 2, 1) &\dec    5.000000000$\times 10^{-1}$ &( 6, 3) &\dec    7.812500000$\times 10^{-2}$ &(10,13) &\dec    2.725607152$\times 10^{-3}$ &(17,12) &\dec $-$8.525536617$\times 10^{-3}$ \\
 ( 2, 2) &\dec    6.250000000$\times 10^{-2}$ &( 6, 4) &\dec    7.595486111$\times 10^{-2}$ &(11, 5) &\dec $-$7.690429688$\times 10^{-3}$ &(17,13) &\dec $-$8.551663397$\times 10^{-3}$ \\
 ( 2, 3) &\dec $-$3.645833333$\times 10^{-2}$ &( 6, 5) &\dec    3.919813368$\times 10^{-2}$ &(11, 6) &\dec $-$1.887210799$\times 10^{-2}$ &(18, 9) &\dec    7.244944572$\times 10^{-4}$ \\
 ( 2, 4) &\dec $-$1.443142361$\times 10^{-2}$ &( 6, 6) &\dec    1.485306540$\times 10^{-2}$ &(11, 7) &\dec $-$2.397200110$\times 10^{-2}$ &(18,10) &\dec    2.481638461$\times 10^{-3}$ \\
 ( 2, 5) &\dec $-$3.861038773$\times 10^{-3}$ &( 6, 7) &\dec    5.066873213$\times 10^{-3}$ &(11, 8) &\dec $-$2.190082988$\times 10^{-2}$ &(18,11) &\dec    4.609868075$\times 10^{-3}$ \\
 ( 2, 6) &\dec $-$4.464373176$\times 10^{-3}$ &( 6, 8) &\dec    1.200659217$\times 10^{-3}$ &(11, 9) &\dec $-$1.703925617$\times 10^{-2}$ &(18,12) &\dec    6.277698444$\times 10^{-3}$ \\
 ( 2, 7) &\dec $-$3.574776355$\times 10^{-3}$ &( 6, 9) &\dec $-$4.880865156$\times 10^{-4}$ &(11,10) &\dec $-$1.243639951$\times 10^{-2}$ &(18,13) &\dec    7.142219966$\times 10^{-3}$ \\
 ( 2, 8) &\dec $-$2.152545002$\times 10^{-3}$ &( 6,10) &\dec $-$1.061951430$\times 10^{-3}$ &(11,11) &\dec $-$8.818602901$\times 10^{-3}$ &(19, 9) &\dec $-$3.622472286$\times 10^{-4}$ \\
 ( 2, 9) &\dec $-$1.614262178$\times 10^{-3}$ &( 6,11) &\dec $-$1.168436483$\times 10^{-3}$ &(11,12) &\dec $-$6.103055681$\times 10^{-3}$ &(19,10) &\dec $-$1.483667264$\times 10^{-3}$ \\
 ( 2,10) &\dec $-$1.364268756$\times 10^{-3}$ &( 6,12) &\dec $-$1.145143702$\times 10^{-3}$ &(11,13) &\dec $-$4.131112380$\times 10^{-3}$ &(19,11) &\dec $-$3.187079773$\times 10^{-3}$ \\
 ( 2,11) &\dec $-$1.085128141$\times 10^{-3}$ &( 6,13) &\dec $-$1.075193082$\times 10^{-3}$ &(12, 6) &\dec    7.049560547$\times 10^{-3}$ &(19,12) &\dec $-$4.863723272$\times 10^{-3}$ \\
 ( 2,12) &\dec $-$8.633290964$\times 10^{-4}$ &( 7, 3) &\dec $-$3.906250000$\times 10^{-2}$ &(12, 7) &\dec    1.552691676$\times 10^{-2}$ &(19,13) &\dec $-$6.030015201$\times 10^{-3}$ \\
 ( 2,13) &\dec $-$7.178656848$\times 10^{-4}$ &( 7, 4) &\dec $-$6.331380208$\times 10^{-2}$ &(12, 8) &\dec    1.896538987$\times 10^{-2}$ &(20,10) &\dec    3.441348672$\times 10^{-4}$ \\
 ( 3, 1) &\dec $-$2.500000000$\times 10^{-1}$ &( 7, 5) &\dec $-$4.956506800$\times 10^{-2}$ &(12, 9) &\dec    1.764401420$\times 10^{-2}$ &(20,11) &\dec    1.318864692$\times 10^{-3}$ \\
 ( 3, 2) &\dec $-$1.875000000$\times 10^{-1}$ &( 7, 6) &\dec $-$2.639009923$\times 10^{-2}$ &(12,10) &\dec    1.446797193$\times 10^{-2}$ &(20,12) &\dec    2.724516711$\times 10^{-3}$ \\
 ( 3, 3) &\dec $-$2.343750000$\times 10^{-2}$ &( 7, 7) &\dec $-$1.199321708$\times 10^{-2}$ &(12,11) &\dec    1.121713674$\times 10^{-2}$ &(20,13) &\dec    4.090782507$\times 10^{-3}$ \\
 ( 3, 4) &\dec    2.012803819$\times 10^{-2}$ &( 7, 8) &\dec $-$5.185196227$\times 10^{-3}$ &(12,12) &\dec    8.435888979$\times 10^{-3}$ &(21,10) &\dec $-$1.720674336$\times 10^{-4}$ \\
 ( 3, 5) &\dec    1.035789207$\times 10^{-2}$ &( 7, 9) &\dec $-$1.776089397$\times 10^{-3}$ &(12,13) &\dec    6.212068934$\times 10^{-3}$ &(21,11) &\dec $-$7.750395765$\times 10^{-4}$ \\
 ( 3, 6) &\dec    4.805176346$\times 10^{-3}$ &( 7,10) &\dec $-$8.192333266$\times 10^{-5}$ &(13, 6) &\dec $-$3.524780273$\times 10^{-3}$ &(21,12) &\dec $-$1.829266700$\times 10^{-3}$ \\
 ( 3, 7) &\dec    4.284713101$\times 10^{-3}$ &( 7,11) &\dec    6.216205777$\times 10^{-4}$ &(13, 7) &\dec $-$1.010268412$\times 10^{-2}$ &(21,13) &\dec $-$3.055890338$\times 10^{-3}$ \\
 ( 3, 8) &\dec    3.284367466$\times 10^{-3}$ &( 7,12) &\dec    8.737107987$\times 10^{-4}$ &(13, 8) &\dec $-$1.511986766$\times 10^{-2}$ &(22,11) &\dec    1.642461866$\times 10^{-4}$ \\
 ( 3, 9) &\dec    2.260859277$\times 10^{-3}$ &( 7,13) &\dec    9.485913165$\times 10^{-4}$ &(13, 9) &\dec $-$1.626035314$\times 10^{-2}$ &(22,12) &\dec    6.962934102$\times 10^{-4}$ \\
 ( 3,10) &\dec    1.762829643$\times 10^{-3}$ &( 8, 4) &\dec    3.417968750$\times 10^{-2}$ &(13,10) &\dec $-$1.465683078$\times 10^{-2}$ &(22,13) &\dec    1.582696489$\times 10^{-3}$ \\
 ( 3,11) &\dec    1.451718673$\times 10^{-3}$ &( 8, 5) &\dec    4.730902778$\times 10^{-2}$ &(13,11) &\dec $-$1.210970065$\times 10^{-2}$ &(23,11) &\dec $-$8.212309331$\times 10^{-5}$ \\
 ( 3,12) &\dec    1.173179929$\times 10^{-3}$ &( 8, 6) &\dec    3.645070394$\times 10^{-2}$ &(13,12) &\dec $-$9.584599362$\times 10^{-3}$ &(23,12) &\dec $-$4.034250937$\times 10^{-4}$ \\
 ( 3,13) &\dec    9.595019662$\times 10^{-4}$ &( 8, 7) &\dec    2.204519358$\times 10^{-2}$ &(13,13) &\dec $-$7.395426977$\times 10^{-3}$ &(23,13) &\dec $-$1.037148415$\times 10^{-3}$ \\
 ( 4, 2) &\dec    1.875000000$\times 10^{-1}$ &( 8, 8) &\dec    1.249375691$\times 10^{-2}$ &(14, 7) &\dec    3.273010254$\times 10^{-3}$ &(24,12) &\dec    7.870129775$\times 10^{-5}$ \\
 ( 4, 3) &\dec    1.041666667$\times 10^{-1}$ &( 8, 9) &\dec    6.866897220$\times 10^{-3}$ &(14, 8) &\dec    8.544238997$\times 10^{-3}$ &(24,13) &\dec    3.656570992$\times 10^{-4}$ \\
 ( 4, 4) &\dec    1.909722222$\times 10^{-2}$ &( 8,10) &\dec    3.496121257$\times 10^{-3}$ &(14, 9) &\dec    1.227229300$\times 10^{-2}$ &(25,12) &\dec $-$3.935064888$\times 10^{-5}$ \\
 ( 4, 5) &\dec $-$3.906250000$\times 10^{-3}$ &( 8,11) &\dec    1.579388581$\times 10^{-3}$ &(14,10) &\dec    1.322238413$\times 10^{-2}$ &(25,13) &\dec $-$2.093578942$\times 10^{-4}$ \\
 ( 4, 6) &\dec $-$3.848888256$\times 10^{-3}$ &( 8,12) &\dec    5.538137269$\times 10^{-4}$ &(14,11) &\dec    1.227668397$\times 10^{-2}$ &(26,13) &\dec    3.783716238$\times 10^{-5}$ \\
 ( 4, 7) &\dec $-$3.145002043$\times 10^{-3}$ &( 8,13) &\dec    4.738508965$\times 10^{-6}$ &(14,12) &\dec    1.055795986$\times 10^{-2}$ &(27,13) &\dec $-$1.891858119$\times 10^{-5}$ \\
 ( 4, 8) &\dec $-$2.985769420$\times 10^{-3}$ &( 9, 4) &\dec $-$1.708984375$\times 10^{-2}$ &(14,13) &\dec    8.705553180$\times 10^{-3}$ & \\
 ( 4, 9) &\dec $-$2.431829046$\times 10^{-3}$ &( 9, 5) &\dec $-$3.485333478$\times 10^{-2}$ &(15, 7) &\dec $-$1.636505127$\times 10^{-3}$ & \\
 ( 4,10) &\dec $-$1.912181882$\times 10^{-3}$ &( 9, 6) &\dec $-$3.606061582$\times 10^{-2}$ &(15, 8) &\dec $-$5.362708709$\times 10^{-3}$ & \\
\end{tabular}
\end{table}

\newpage
\begin{table}
\squeezetable
\setdec 0.00000000000000
\caption{Series coefficients for
the exclusive structure factors of the 1-particle triplet state $S_{\rm 1p}(k)$
$ S_{\rm 1p}(k)$ $ = \sum_{n,m} a_{n,m}$ $ \lambda^m \cos (n kd/2) $.
Nonzero coefficients $a_{n,m}$
up to order $m=13$ are listed.
  }
\label{tab2}
\begin{tabular}{|rr|rr|rr|rr|}
 \multicolumn{1}{|c}{$(n,m)$} &\multicolumn{1}{c|}{$a_{n,m}$}
& \multicolumn{1}{c}{$(n,m)$} &\multicolumn{1}{c|}{$a_{n,m}$}
& \multicolumn{1}{c}{$(n,m)$} &\multicolumn{1}{c|}{$a_{n,m}$}
& \multicolumn{1}{c}{$(n,m)$} &\multicolumn{1}{c|}{$a_{n,m}$} \\
\tableline
 ( 0, 0) &\dec    1.000000000                 &( 4, 2) &\dec    1.875000000$\times 10^{-1}$ &( 8,11) &\dec $-$6.011492935$\times 10^{-3}$ &(14,13) &\dec    8.674883160$\times 10^{-3}$ \\
 ( 0, 2) &\dec $-$3.125000000$\times 10^{-1}$ &( 4, 3) &\dec    1.458333333$\times 10^{-1}$ &( 8,12) &\dec    9.711804995$\times 10^{-3}$ &(15, 7) &\dec $-$1.636505127$\times 10^{-3}$ \\
 ( 0, 3) &\dec $-$9.375000000$\times 10^{-2}$ &( 4, 4) &\dec    6.835937500$\times 10^{-3}$ &( 8,13) &\dec    4.710318350$\times 10^{-3}$ &(15, 8) &\dec $-$5.358020138$\times 10^{-3}$ \\
 ( 0, 4) &\dec    1.627604167$\times 10^{-2}$ &( 4, 5) &\dec $-$5.485930266$\times 10^{-2}$ &( 9, 4) &\dec $-$1.708984375$\times 10^{-2}$ &(15, 9) &\dec $-$9.241334017$\times 10^{-3}$ \\
 ( 0, 5) &\dec $-$8.257378472$\times 10^{-2}$ &( 4, 6) &\dec    9.334422924$\times 10^{-3}$ &( 9, 5) &\dec $-$3.543203848$\times 10^{-2}$ &(15,10) &\dec $-$1.131045012$\times 10^{-2}$ \\
 ( 0, 6) &\dec $-$4.077148438$\times 10^{-2}$ &( 4, 7) &\dec    7.197318057$\times 10^{-3}$ &( 9, 6) &\dec $-$3.456574899$\times 10^{-2}$ &(15,11) &\dec $-$1.142877596$\times 10^{-2}$ \\
 ( 0, 7) &\dec    1.841892038$\times 10^{-2}$ &( 4, 8) &\dec $-$5.832889946$\times 10^{-2}$ &( 9, 7) &\dec $-$2.518777141$\times 10^{-2}$ &(15,12) &\dec $-$1.046099406$\times 10^{-2}$ \\
 ( 0, 8) &\dec $-$4.383319893$\times 10^{-2}$ &( 4, 9) &\dec $-$1.281724218$\times 10^{-2}$ &( 9, 8) &\dec $-$1.987998380$\times 10^{-2}$ &(15,13) &\dec $-$9.327327593$\times 10^{-3}$ \\
 ( 0, 9) &\dec $-$3.392954635$\times 10^{-2}$ &( 4,10) &\dec    2.152346878$\times 10^{-2}$ &( 9, 9) &\dec $-$1.377372682$\times 10^{-2}$ &(16, 8) &\dec    1.534223557$\times 10^{-3}$ \\
 ( 0,10) &\dec    1.671238124$\times 10^{-2}$ &( 4,11) &\dec $-$4.067856177$\times 10^{-2}$ &( 9,10) &\dec $-$1.514186704$\times 10^{-3}$ &(16, 9) &\dec    4.632624525$\times 10^{-3}$ \\
 ( 0,11) &\dec $-$2.402930407$\times 10^{-2}$ &( 4,12) &\dec $-$2.401466656$\times 10^{-2}$ &( 9,11) &\dec $-$2.640524172$\times 10^{-4}$ &(16,10) &\dec    7.616553597$\times 10^{-3}$ \\
 ( 0,12) &\dec $-$3.094915187$\times 10^{-2}$ &( 4,13) &\dec    2.523955463$\times 10^{-2}$ &( 9,12) &\dec $-$8.667418864$\times 10^{-3}$ &(16,11) &\dec    9.310862860$\times 10^{-3}$ \\
 ( 0,13) &\dec    1.317777693$\times 10^{-2}$ &( 5, 2) &\dec $-$9.375000000$\times 10^{-2}$ &( 9,13) &\dec $-$3.106212436$\times 10^{-4}$ &(16,12) &\dec    9.677973448$\times 10^{-3}$ \\
 ( 1, 0) &\dec $-$1.000000000                 &( 5, 3) &\dec $-$1.223958333$\times 10^{-1}$ &(10, 5) &\dec    1.538085938$\times 10^{-2}$ &(16,13) &\dec    9.174113955$\times 10^{-3}$ \\
 ( 1, 1) &\dec $-$2.500000000$\times 10^{-1}$ &( 5, 4) &\dec $-$4.031032986$\times 10^{-2}$ &(10, 6) &\dec    2.725408107$\times 10^{-2}$ &(17, 8) &\dec $-$7.671117783$\times 10^{-4}$ \\
 ( 1, 2) &\dec    5.000000000$\times 10^{-1}$ &( 5, 5) &\dec $-$1.652470341$\times 10^{-3}$ &(10, 7) &\dec    2.832565779$\times 10^{-2}$ &(17, 9) &\dec $-$2.828986461$\times 10^{-3}$ \\
 ( 1, 3) &\dec    1.432291667$\times 10^{-1}$ &( 5, 6) &\dec $-$2.680371131$\times 10^{-2}$ &(10, 8) &\dec    2.311140563$\times 10^{-2}$ &(17,10) &\dec $-$5.461641059$\times 10^{-3}$ \\
 ( 1, 4) &\dec $-$3.797743056$\times 10^{-2}$ &( 5, 7) &\dec $-$2.120727570$\times 10^{-3}$ &(10, 9) &\dec    1.449731923$\times 10^{-2}$ &(17,11) &\dec $-$7.554290412$\times 10^{-3}$ \\
 ( 1, 5) &\dec    1.475604022$\times 10^{-1}$ &( 5, 8) &\dec    3.677335722$\times 10^{-2}$ &(10,10) &\dec    6.246098908$\times 10^{-3}$ &(17,12) &\dec $-$8.549734505$\times 10^{-3}$ \\
 ( 1, 6) &\dec    6.164414206$\times 10^{-2}$ &( 5, 9) &\dec $-$4.177481543$\times 10^{-3}$ &(10,11) &\dec    6.405727927$\times 10^{-3}$ &(17,13) &\dec $-$8.565778023$\times 10^{-3}$ \\
 ( 1, 7) &\dec $-$4.571135603$\times 10^{-2}$ &( 5,10) &\dec $-$1.510927532$\times 10^{-2}$ &(10,12) &\dec    8.071629661$\times 10^{-3}$ &(18, 9) &\dec    7.244944572$\times 10^{-4}$ \\
 ( 1, 8) &\dec    7.541900211$\times 10^{-2}$ &( 5,11) &\dec    3.391969899$\times 10^{-2}$ &(10,13) &\dec $-$1.769051814$\times 10^{-4}$ &(18,10) &\dec    2.481229141$\times 10^{-3}$ \\
 ( 1, 9) &\dec    5.347557773$\times 10^{-2}$ &( 5,12) &\dec    1.090426758$\times 10^{-2}$ &(11, 5) &\dec $-$7.690429688$\times 10^{-3}$ &(18,11) &\dec    4.613304272$\times 10^{-3}$ \\
 ( 1,10) &\dec $-$4.190407427$\times 10^{-2}$ &( 5,13) &\dec $-$2.101789738$\times 10^{-2}$ &(11, 6) &\dec $-$1.875154472$\times 10^{-2}$ &(18,12) &\dec    6.275948659$\times 10^{-3}$ \\
 ( 1,11) &\dec    3.910767511$\times 10^{-2}$ &( 6, 3) &\dec    7.812500000$\times 10^{-2}$ &(11, 7) &\dec $-$2.441788701$\times 10^{-2}$ &(18,13) &\dec    7.129091829$\times 10^{-3}$ \\
 ( 1,12) &\dec    5.118486369$\times 10^{-2}$ &( 6, 4) &\dec    6.727430556$\times 10^{-2}$ &(11, 8) &\dec $-$2.218640450$\times 10^{-2}$ &(19, 9) &\dec $-$3.622472286$\times 10^{-4}$ \\
 ( 1,13) &\dec $-$3.356079789$\times 10^{-2}$ &( 6, 5) &\dec    4.458279080$\times 10^{-2}$ &(11, 9) &\dec $-$1.605467746$\times 10^{-2}$ &(19,10) &\dec $-$1.483499815$\times 10^{-3}$ \\
 ( 2, 1) &\dec    5.000000000$\times 10^{-1}$ &( 6, 6) &\dec    3.504162070$\times 10^{-2}$ &(11,10) &\dec $-$1.111541180$\times 10^{-2}$ &(19,11) &\dec $-$3.188966289$\times 10^{-3}$ \\
 ( 2, 2) &\dec $-$1.250000000$\times 10^{-1}$ &( 6, 7) &\dec    2.109803094$\times 10^{-3}$ &(11,11) &\dec $-$1.034165700$\times 10^{-2}$ &(19,12) &\dec $-$4.859982468$\times 10^{-3}$ \\
 ( 2, 3) &\dec $-$2.604166667$\times 10^{-2}$ &( 6, 8) &\dec $-$1.579822777$\times 10^{-2}$ &(11,12) &\dec $-$8.030332288$\times 10^{-3}$ &(19,13) &\dec $-$6.025154806$\times 10^{-3}$ \\
 ( 2, 4) &\dec    1.779513889$\times 10^{-2}$ &( 6, 9) &\dec    1.297988287$\times 10^{-2}$ &(11,13) &\dec $-$1.448277762$\times 10^{-3}$ &(20,10) &\dec    3.441348672$\times 10^{-4}$ \\
 ( 2, 5) &\dec $-$1.206416377$\times 10^{-1}$ &( 6,10) &\dec    6.764275332$\times 10^{-3}$ &(12, 6) &\dec    7.049560547$\times 10^{-3}$ &(20,11) &\dec    1.318939114$\times 10^{-3}$ \\
 ( 2, 6) &\dec $-$2.833321654$\times 10^{-2}$ &( 6,11) &\dec $-$2.652577291$\times 10^{-2}$ &(12, 7) &\dec    1.559121717$\times 10^{-2}$ &(20,12) &\dec    2.723720941$\times 10^{-3}$ \\
 ( 2, 7) &\dec    4.500562644$\times 10^{-2}$ &( 6,12) &\dec $-$1.873115433$\times 10^{-3}$ &(12, 8) &\dec    1.875765450$\times 10^{-2}$ &(20,13) &\dec    4.091885849$\times 10^{-3}$ \\
 ( 2, 8) &\dec $-$6.561594063$\times 10^{-2}$ &( 6,13) &\dec    1.362791590$\times 10^{-2}$ &(12, 9) &\dec    1.728643232$\times 10^{-2}$ &(21,10) &\dec $-$1.720674336$\times 10^{-4}$ \\
 ( 2, 9) &\dec $-$3.427604879$\times 10^{-2}$ &( 7, 3) &\dec $-$3.906250000$\times 10^{-2}$ &(12,10) &\dec    1.481153271$\times 10^{-2}$ &(21,11) &\dec $-$7.750705856$\times 10^{-4}$ \\
 ( 2,10) &\dec    4.545736434$\times 10^{-2}$ &( 7, 4) &\dec $-$6.070963542$\times 10^{-2}$ &(12,11) &\dec    1.267199966$\times 10^{-2}$ &(21,12) &\dec $-$1.828835480$\times 10^{-3}$ \\
 ( 2,11) &\dec $-$3.409665942$\times 10^{-2}$ &( 7, 5) &\dec $-$5.429642289$\times 10^{-2}$ &(12,12) &\dec    8.696286848$\times 10^{-3}$ &(21,13) &\dec $-$3.057141912$\times 10^{-3}$ \\
 ( 2,12) &\dec $-$3.879019409$\times 10^{-2}$ &( 7, 6) &\dec $-$3.120500070$\times 10^{-2}$ &(12,13) &\dec    4.298281913$\times 10^{-3}$ &(22,11) &\dec    1.642461866$\times 10^{-4}$ \\
 ( 2,13) &\dec    3.772049749$\times 10^{-2}$ &( 7, 7) &\dec $-$3.188448384$\times 10^{-3}$ &(13, 6) &\dec $-$3.524780273$\times 10^{-3}$ &(22,12) &\dec    6.962799729$\times 10^{-4}$ \\
 ( 3, 1) &\dec $-$2.500000000$\times 10^{-1}$ &( 7, 8) &\dec    1.279329754$\times 10^{-3}$ &(13, 7) &\dec $-$1.012679677$\times 10^{-2}$ &(22,13) &\dec    1.582874937$\times 10^{-3}$ \\
 ( 3, 2) &\dec $-$1.562500000$\times 10^{-1}$ &( 7, 9) &\dec $-$1.467444993$\times 10^{-2}$ &(13, 8) &\dec $-$1.499607686$\times 10^{-2}$ &(23,11) &\dec $-$8.212309331$\times 10^{-5}$ \\
 ( 3, 3) &\dec $-$8.593750000$\times 10^{-2}$ &( 7,10) &\dec $-$9.349685042$\times 10^{-4}$ &(13, 9) &\dec $-$1.624052105$\times 10^{-2}$ &(23,12) &\dec $-$4.034194087$\times 10^{-4}$ \\
 ( 3, 4) &\dec    1.372612847$\times 10^{-2}$ &( 7,11) &\dec    1.714891750$\times 10^{-2}$ &(13,10) &\dec $-$1.498198003$\times 10^{-2}$ &(23,13) &\dec $-$1.037244158$\times 10^{-3}$ \\
 ( 3, 5) &\dec    1.005768953$\times 10^{-1}$ &( 7,12) &\dec $-$5.294607061$\times 10^{-3}$ &(13,11) &\dec $-$1.252169240$\times 10^{-2}$ &(24,12) &\dec    7.870129775$\times 10^{-5}$ \\
 ( 3, 6) &\dec    9.286244710$\times 10^{-3}$ &( 7,13) &\dec $-$8.581846951$\times 10^{-3}$ &(13,12) &\dec $-$9.049303257$\times 10^{-3}$ &(24,13) &\dec    3.656595110$\times 10^{-4}$ \\
 ( 3, 7) &\dec $-$2.312368150$\times 10^{-2}$ &( 8, 4) &\dec    3.417968750$\times 10^{-2}$ &(13,13) &\dec $-$6.457204366$\times 10^{-3}$ &(25,12) &\dec $-$3.935064888$\times 10^{-5}$ \\
 ( 3, 8) &\dec    6.813065057$\times 10^{-2}$ &( 8, 5) &\dec    4.904513889$\times 10^{-2}$ &(14, 7) &\dec    3.273010254$\times 10^{-3}$ &(25,13) &\dec $-$2.093589278$\times 10^{-4}$ \\
 ( 3, 9) &\dec    2.664720824$\times 10^{-2}$ &( 8, 6) &\dec    3.434541490$\times 10^{-2}$ &(14, 8) &\dec    8.532182670$\times 10^{-3}$ &(26,13) &\dec    3.783716238$\times 10^{-5}$ \\
 ( 3,10) &\dec $-$3.218290488$\times 10^{-2}$ &( 8, 7) &\dec    1.559162061$\times 10^{-2}$ &(14, 9) &\dec    1.232853596$\times 10^{-2}$ &(27,13) &\dec $-$1.891858119$\times 10^{-5}$ \\
 ( 3,11) &\dec    4.070336366$\times 10^{-2}$ &( 8, 8) &\dec    1.322605786$\times 10^{-2}$ &(14,10) &\dec    1.328042130$\times 10^{-2}$ &        &                                    \\
 ( 3,12) &\dec    3.466875907$\times 10^{-2}$ &( 8, 9) &\dec    1.580418652$\times 10^{-2}$ &(14,11) &\dec    1.213368398$\times 10^{-2}$ &        &                                    \\
 ( 3,13) &\dec $-$3.003621683$\times 10^{-2}$ &( 8,10) &\dec    9.329997203$\times 10^{-4}$ &(14,12) &\dec    1.012086983$\times 10^{-2}$ &        &                                    \\
\end{tabular}
\end{table}

\newpage
\begin{table}
\squeezetable
\setdec 0.00000000000000
\caption{Series coefficients for the integrated structure factor $S(k)$,
the exclusive 1-particle structure factor $S_{\rm 1p}(k)$, and
the total 2-particle structure factor $S_{\rm 2p}(k)$ at
$kd=\pi$, $2\pi$, together with the structure factors
($S_{T_1}(k)$, $S_{T_2}(k)$, and $S_{\rm 2pc}(k)$) for 2-particle
bound states $T_1$ and $T_2$, and 2-particle continuum at $kd=\pi$,
  the auto correlation function
of 1-particle ($\Phi_{\rm 1p}$) and 2-particle states ($\Phi_{\rm 2p}$),
the average relative weight for  2-particle states ($\overline{W}_{\rm 2p}$),
and $R$, $R_{\rm 1p}$, and $R_{\rm 2p}$.
Series coefficients of $\lambda^m$
up to order $m=13$ are listed.
}\label{tab4}
\begin{tabular}{|rrrrrr|}
  \multicolumn{1}{|c}{$m$}
& \multicolumn{1}{c}{$S(\pi )$}     & \multicolumn{1}{c}{$S_{\rm 1p}(\pi)$}
& \multicolumn{1}{c}{$S_{\rm 2p}(\pi)$}  & \multicolumn{1}{c}{$S_{T_1}(\pi)$}
& \multicolumn{1}{c|}{$S_{T_2}(\pi)$}   \\
\tableline
%             S(\pi)                            S_{\rm 1p}(\pi)                         S_{\rm 2p})(\pi)                     S_{T_1}(\pi)                        S_{T_2}(\pi)
  0 &\dec    1.000000000                 &\dec    1.000000000                 &\dec    0.000000000                 &\dec    0.000000000                 &\dec    0.000000000                 \\
  1 &\dec $-$5.000000000$\times 10^{-1}$ &\dec $-$5.000000000$\times 10^{-1}$ &\dec    0.000000000                 &\dec    0.000000000                 &\dec    0.000000000                 \\
  2 &\dec    1.250000000$\times 10^{-1}$ &\dec    0.000000000                 &\dec    1.250000000$\times 10^{-1}$ &\dec    1.250000000$\times 10^{-1}$ &\dec    0.000000000                 \\
  3 &\dec    6.250000000$\times 10^{-2}$ &\dec    0.000000000                 &\dec    6.250000000$\times 10^{-2}$ &\dec    6.250000000$\times 10^{-2}$ &\dec    0.000000000                 \\
  4 &\dec $-$8.246527778$\times 10^{-3}$ &\dec $-$2.777777778$\times 10^{-2}$ &\dec    1.584201389$\times 10^{-2}$ &\dec    4.557291667$\times 10^{-3}$ &\dec    2.604166667$\times 10^{-3}$ \\
  5 &\dec $-$7.315176505$\times 10^{-3}$ &\dec $-$2.770996094$\times 10^{-2}$ &\dec    1.725260417$\times 10^{-2}$ &\dec $-$2.365451389$\times 10^{-2}$ &\dec    5.642361111$\times 10^{-3}$ \\
  6 &\dec    1.671025782$\times 10^{-3}$ &\dec $-$2.400457123$\times 10^{-2}$ &\dec    2.357124988$\times 10^{-2}$ &\dec $-$6.690809462$\times 10^{-2}$ &\dec    4.643192998$\times 10^{-3}$ \\
  7 &\dec    2.038836381$\times 10^{-3}$ &\dec $-$2.191502135$\times 10^{-2}$ &\dec    2.248861054$\times 10^{-2}$ &\dec $-$1.295114399$\times 10^{-1}$ &\dec $-$3.105917095$\times 10^{-3}$ \\
  8 &\dec    9.809032520$\times 10^{-4}$ &\dec $-$1.887358236$\times 10^{-2}$ &\dec    1.753652230$\times 10^{-2}$ &\dec $-$1.919116349$\times 10^{-1}$ &\dec $-$1.721538889$\times 10^{-2}$ \\
  9 &\dec    7.194422822$\times 10^{-4}$ &\dec $-$1.527772889$\times 10^{-2}$ &\dec    1.378672204$\times 10^{-2}$ &\dec $-$2.252182500$\times 10^{-1}$ &\dec $-$3.106542706$\times 10^{-2}$ \\
 10 &\dec    4.813229257$\times 10^{-4}$ &\dec $-$1.228831810$\times 10^{-2}$ &\dec    1.083241128$\times 10^{-2}$ &\dec $-$1.979558323$\times 10^{-1}$ &\dec $-$3.789373385$\times 10^{-2}$ \\
 11 &\dec    2.551816973$\times 10^{-4}$ &\dec $-$1.011208718$\times 10^{-2}$ &\dec    9.645537754$\times 10^{-3}$ &\dec $-$7.313449239$\times 10^{-2}$ &\dec $-$3.170664884$\times 10^{-2}$ \\
 12 &\dec    1.953605167$\times 10^{-4}$ &\dec $-$8.576749497$\times 10^{-3}$ &\dec    7.501091661$\times 10^{-3}$ &\dec    1.758607061$\times 10^{-1}$ &\dec $-$6.069523293$\times 10^{-3}$ \\
 13 &\dec    2.020270912$\times 10^{-4}$ &\dec $-$7.538604157$\times 10^{-3}$ &\dec    7.444260493$\times 10^{-3}$ &\dec    5.577962015$\times 10^{-1}$ &\dec    4.352098881$\times 10^{-2}$ \\
 14 &                                    &                                    &                                    &                                    &\dec    1.144814294$\times 10^{-1}$ \\
\tableline
\tableline
 \multicolumn{1}{|c}{$m$} & \multicolumn{1}{c}{$S_{\rm 2pc}(\pi)$}
& \multicolumn{1}{c}{$S(2\pi )$}     & \multicolumn{1}{c}{$S_{\rm 1p}(2\pi)$}
& \multicolumn{1}{c}{$S_{\rm 2p}(2\pi)$}  & \multicolumn{1}{c|}{$\Phi_{\rm 1p}$}   \\
\tableline
%            S_{\rm 2pc}(\pi)                         S(2\pi)                           S_{\rm 1p}(2\pi)                         S_{\rm 2p})(2\pi)                       \Phi_{\rm 1p})
  0 &\dec    0.000000000                 &\dec    2.000000000                 &\dec    2.000000000                 &\dec    0.000000000                 &\dec    1.000000000       \\
  1 &\dec    0.000000000                 &\dec    1.000000000                 &\dec    1.000000000                 &\dec    0.000000000                 &\dec    0.000000000       \\
  2 &\dec    0.000000000                 &\dec    5.000000000$\times 10^{-1}$ &\dec $-$5.000000000$\times 10^{-1}$ &\dec    1.000000000                 &\dec $-$3.125000000$\times 10^{-1}$ \\
  3 &\dec    0.000000000                 &\dec    2.916666667$\times 10^{-1}$ &\dec    2.083333333$\times 10^{-1}$ &\dec    8.333333333$\times 10^{-2}$ &\dec $-$9.375000000$\times 10^{-2}$ \\
  4 &\dec    8.680555556$\times 10^{-3}$ &\dec    2.296006944$\times 10^{-1}$ &\dec    2.847222222$\times 10^{-1}$ &\dec $-$9.027777778$\times 10^{-2}$ &\dec    1.627604167$\times 10^{-2}$ \\
  5 &\dec    3.526475694$\times 10^{-2}$ &\dec    1.882414641$\times 10^{-1}$ &\dec $-$2.981318721$\times 10^{-1}$ &\dec    4.924768519$\times 10^{-1}$ &\dec $-$8.257378472$\times 10^{-2}$ \\
  6 &\dec    8.583615150$\times 10^{-2}$ &\dec    1.552634534$\times 10^{-1}$ &\dec    8.784079846$\times 10^{-2}$ &\dec    1.213981723$\times 10^{-1}$ &\dec $-$4.077148438$\times 10^{-2}$ \\
  7 &\dec    1.551059676$\times 10^{-1}$ &\dec    1.336307604$\times 10^{-1}$ &\dec    2.710263476$\times 10^{-1}$ &\dec $-$1.130861384$\times 10^{-1}$ &\dec    1.841892038$\times 10^{-2}$ \\
  8 &\dec    2.266635461$\times 10^{-1}$ &\dec    1.180685972$\times 10^{-1}$ &\dec $-$2.368294851$\times 10^{-1}$ &\dec    3.319204641$\times 10^{-1}$ &\dec $-$4.383319893$\times 10^{-2}$ \\
  9 &\dec    2.700703991$\times 10^{-1}$ &\dec    1.053990124$\times 10^{-1}$ &\dec $-$5.538722908$\times 10^{-3}$ &\dec    2.095799705$\times 10^{-1}$ &\dec $-$3.392954635$\times 10^{-2}$ \\
 10 &\dec    2.466819774$\times 10^{-1}$ &\dec    9.514478385$\times 10^{-2}$ &\dec    2.723409199$\times 10^{-1}$ &\dec    3.264975567$\times 10^{-2}$ &\dec    1.671238124$\times 10^{-2}$ \\
 11 &\dec    1.144866790$\times 10^{-1}$ &\dec    8.684667562$\times 10^{-2}$ &\dec $-$1.694460542$\times 10^{-1}$ &\dec    4.627752009$\times 10^{-1}$ &\dec $-$2.402930407$\times 10^{-2}$ \\
 12 &\dec $-$1.622900912$\times 10^{-1}$ &\dec    7.991574506$\times 10^{-2}$ &\dec $-$7.914782458$\times 10^{-2}$ &\dec    4.466072907$\times 10^{-1}$ &\dec $-$3.094915187$\times 10^{-2}$ \\
 13 &\dec $-$5.938729299$\times 10^{-1}$ &\dec    7.400747491$\times 10^{-2}$ &\dec    2.593075729$\times 10^{-1}$ &\dec    7.243209793$\times 10^{-1}$ &\dec    1.317777693$\times 10^{-2}$ \\
\tableline
\tableline
\multicolumn{1}{|c}{$m$}
 & \multicolumn{1}{c}{$\Phi_{\rm 2p}$} & \multicolumn{1}{c}{$\overline{W}_{\rm 2p}$}
 & \multicolumn{1}{c}{$R=\lim_{k\to 0} S/k^2$}
 & \multicolumn{1}{c}{$R_{\rm 1p}=\lim_{k\to 0} S_{\rm 1p}/k^2$}
 & \multicolumn{1}{c|}{$R_{\rm 2p}=\lim_{k\to 0} S_{\rm 2p}/k^2$} \\
\tableline
%             \Phi_{\rm 2p}                      average W_{\rm 2p}                          R                                    R_{\rm 1p}                               R_{\rm 2p}
  0 &\dec    0.000000000                 &\dec    1.000000000                  &\dec    1.250000000$\times 10^{-1}$ &\dec    1.250000000$\times 10^{-1}$ &\dec    0.000000000                  \\
  1 &\dec    0.000000000                 &\dec    0.000000000                  &\dec    6.250000000$\times 10^{-2}$ &\dec    6.250000000$\times 10^{-2}$ &\dec    0.000000000                  \\
  2 &\dec    3.125000000$\times 10^{-1}$ &\dec $-$1.875000000$\times 10^{-1}$  &\dec    9.375000000$\times 10^{-2}$ &\dec    9.375000000$\times 10^{-2}$ &\dec    0.000000000                  \\
  3 &\dec    9.375000000$\times 10^{-2}$ &\dec $-$4.687500000$\times 10^{-2}$  &\dec    7.031250000$\times 10^{-2}$ &\dec    7.031250000$\times 10^{-2}$ &\dec    0.000000000                  \\
  4 &\dec $-$2.636718750$\times 10^{-2}$ &\dec $-$3.103298611$\times 10^{-2}$  &\dec    6.610785590$\times 10^{-2}$ &\dec    6.141493056$\times 10^{-2}$ &\dec    4.448784722$\times 10^{-3}$  \\
  5 &\dec    8.148871528$\times 10^{-2}$ &\dec $-$6.538447627$\times 10^{-2}$  &\dec    6.383938260$\times 10^{-2}$ &\dec    6.599765354$\times 10^{-2}$ &\dec $-$2.305772569$\times 10^{-3}$  \\
  6 &\dec    5.053823966$\times 10^{-2}$ &\dec $-$1.166864089$\times 10^{-2}$  &\dec    6.018220642$\times 10^{-2}$ &\dec    6.027503661$\times 10^{-2}$ &\dec $-$1.646442178$\times 10^{-4}$  \\
  7 &\dec $-$1.360771980$\times 10^{-2}$ &\dec    2.293814357$\times 10^{-3}$  &\dec    5.737893768$\times 10^{-2}$ &\dec    5.616068546$\times 10^{-2}$ &\dec    9.660760071$\times 10^{-4}$  \\
  8 &\dec    3.967389006$\times 10^{-2}$ &\dec $-$3.920166318$\times 10^{-2}$  &\dec    5.536595911$\times 10^{-2}$ &\dec    5.331091464$\times 10^{-2}$ &\dec    1.625177465$\times 10^{-3}$  \\
  9 &\dec    5.235992915$\times 10^{-2}$ &\dec $-$1.075891724$\times 10^{-2}$  &\dec    5.357448621$\times 10^{-2}$ &\dec    5.601825866$\times 10^{-2}$ &\dec $-$3.020253684$\times 10^{-3}$  \\
 10 &\dec    1.907595256$\times 10^{-2}$ &\dec    7.139608316$\times 10^{-3}$  &\dec    5.194887667$\times 10^{-2}$ &\dec    5.101011394$\times 10^{-2}$ &\dec $-$3.106958042$\times 10^{-5}$  \\
 11 &\dec    6.149786123$\times 10^{-2}$ &\dec $-$2.797421962$\times 10^{-2}$  &\dec    5.053076987$\times 10^{-2}$ &\dec    4.803748655$\times 10^{-2}$ &\dec    6.937664269$\times 10^{-4}$  \\
 12 &\dec    7.568512290$\times 10^{-2}$ &\dec $-$1.386098936$\times 10^{-2}$  &\dec    4.927953902$\times 10^{-2}$ &\dec    5.120296763$\times 10^{-2}$ &\dec $-$5.289988311$\times 10^{-3}$  \\
 13 &\dec    1.267526913$\times 10^{-1}$ &\dec    9.053856936$\times 10^{-3}$  &\dec    4.815020988$\times 10^{-2}$ &\dec    4.831429881$\times 10^{-2}$ &\dec $-$6.876775997$\times 10^{-3}$  \\
\end{tabular}
\end{table}

\begin{table}
\squeezetable
\setdec 0.00000000000000
\caption{$[n/m]$ Dlog Pad\'e approximants to the series for
$\partial S^{2/3}/\partial \lambda$ and
$\partial S^{4/5}/\partial \lambda$  at $kd=2\pi$, for
$\Phi_{\rm 1p}$, and for $\partial \Phi_{\rm 2p}/\partial \lambda$. The position of the
singularity (pole), the critical index (U.B.R.) from unbiased approximants,
and the critical index from biased approximants (B.R.) are given.
An asterisk denotes a defective approximant.
}
 \label{tab5}
\begin{tabular}{rccccc}
\multicolumn{1}{c}{n} &\multicolumn{1}{c}{$[(n-2)/n]$}&\multicolumn{1}{c}{$[(n-1)/n]$}
&\multicolumn{1}{c}{$[n/n]$} &\multicolumn{1}{c}{$[(n+1)/n]$}&\multicolumn{1}{c}{$[(n+2)/n]$}
 \\
\multicolumn{1}{c}{} &\multicolumn{1}{c}{pole (N.B.R. , B.R.)}
&\multicolumn{1}{c}{pole (N.B.R., B.R.)}
&\multicolumn{1}{c}{pole (N.B.R., B.R.)}
&\multicolumn{1}{c}{pole (N.B.R., B.R.)}
&\multicolumn{1}{c}{pole (N.B.R., B.R.)} \\
\tableline
\multicolumn{6}{c}{$\partial S(2\pi)^{2/3}/\partial \lambda$} \\
%series for (S(2Pi))^(2/3) S is Total Weight for Kd,ypara=  6.2832D+00  0.0000D+00
n= 1 &                                 &                                 &\dec  0.6244(-0.2599,~-0.7844) &\dec  1.0859(-1.3670,~-0.9980) &\dec  1.0837(-1.3560,~-0.2298) \\
n= 2 &   ~~~~~~~*~~(~~~*~~~~,~-1.0983) & ~~~~~~~~*~(~~*~~~~,~~~*~~~~) &\dec  1.0837(-1.3561,~-0.9488) &  ~~~~~~~*~~(~~*~~~~,~-0.9530) &\dec  0.9694(-0.7956,~-0.9621) \\
n= 3 &   ~~~~~~~*~~(~~~*~~~~,~-1.1246) &\dec 0.9952(-0.9312,~-0.9534) &\dec  0.9895(-0.9048,~-0.9449) &\dec  0.9933(-0.9257,~-0.9802) &\dec  0.9974(-0.9523,~-0.9747) \\
n= 4 & \dec   0.9899(-0.9074,~-0.9629)   &\dec 0.9918(-0.9168,~-0.9876) & ~~~~~~~~*~~(~~*~~~~,~-0.9738) &\dec  0.9981(-0.9582,~-0.9765) &\dec  0.9980(-0.9576,~-0.9300) \\
n= 5 &   ~~~~~~~*~~(~~~*~~~~,~-0.9764) &\dec 0.9970(-0.9483,~-0.9773) &\dec  0.9980(-0.9575,~-0.9821) &\dec  0.9981(-0.9581,~-0.9832) &                                  \\
n= 6 & \dec   0.9981(-0.9586,~-0.9758)   &\dec 0.9993(-0.9724,~-0.9832) &  &  &                                  \\
% \tableline
% \multicolumn{6}{c}{$\partial S(2\pi)^{3/4}/\partial \lambda$} \\
% %series for (S(2Pi))^(3/4) S is Total Weight for Kd,ypara=  6.2832D+00  0.0000D+00
% n= 1 &                                 &                              &\dec  0.6346(-0.2769,~-0.8147) &\dec  1.0807(-1.3672,~-1.0162) & ~~~~~~~~*~~(~~~~*~~~,~~~~~*~~~)\\
% n= 2 &    ~~~~~~*~~(~~~*~~~~,~-1.0832) & ~~~~~*~~(~~~*~~~~,~~~~*~~~~) &\dec  1.0834(-1.3813,~-0.9673) &\dec  1.0807(-1.3674,~-0.9707) &\dec  0.9704(-0.8149,~-0.9792) \\
% n= 3 &    ~~~~~~*~~(~~~*~~~~,~-1.1150) &\dec 0.9963(-0.9534,~-0.9709) &\dec  0.9903(-0.9252,~-0.9650) &\dec  0.9938(-0.9450,~-0.9961) &\dec  0.9977(-0.9710,~-0.9905) \\
% n= 4 &\dec  0.9908(-0.9281,~-0.9797)   &\dec 0.9925(-0.9372,~-1.0037) & ~~~~~~~~*~~(~~~*~~~,~-0.9895) &\dec  0.9984(-0.9765,~-0.9921) &\dec  0.9983(-0.9756,~-0.9696) \\
% n= 5 &    ~~~~~~*~~(~~~*~~~~,~-0.9921) &\dec 0.9973(-0.9669,~-0.9928) &\dec  0.9983(-0.9754,~-0.9969) &\dec  0.9984(-0.9763,~-0.9977) &                                  \\
% n= 6 &\dec  0.9984(-0.9764,~-0.9918)   &\dec 0.9995(-0.9893,~-0.9977) &  &  &                                  \\
\tableline
\multicolumn{6}{c}{$\partial S(2\pi)^{4/5}/\partial \lambda$} \\
%series for (S(2Pi))^(4/5) S is Total Weight for Kd,ypara=  6.2832D+00  0.0000D+00
n= 1 &                                &                              &\dec  0.6406(-0.2873,~-0.8325) &\dec  1.0777(-1.3676,-1.0271) &\dec  1.0833(-1.3962,~~~~~*~~~) \\
n= 2 &  ~~~~~~~*~~(~~~*~~~~,~-1.0775) &\dec  0.8684(-0.6358,-1.2702) &\dec  1.0833(-1.3967,~-0.9784) &\dec  1.0779(-1.3683,-0.9813) &\dec  0.9710(-0.8264,~-0.9895) \\
n= 3 &  ~~~~~~~*~~(~~~*~~~~,~-1.1123) &\dec  0.9969(-0.9668,-0.9814) &\dec  0.9907(-0.9375,~-0.9768) &\dec  0.9941(-0.9565,-1.0057) &\dec  0.9979(-0.9823,~-1.0000) \\
n= 4 &\dec   0.9913(-0.9406,~-0.9898) &\dec  0.9930(-0.9494,-1.0134) & ~~~~~~~~*~~(~~*~~~~,~-0.9990) &\dec  0.9986(-0.9875,-1.0015) &\dec  0.9985(-0.9865,~-0.9861) \\
n= 5 &  ~~~~~~~*~~(~~~*~~~~,~-1.0016) &\dec  0.9975(-0.9782,-1.0021) &\dec  0.9984(-0.9862,~-1.0058) &\dec  0.9985(-0.9872,-1.0065) &                               \\
n= 6 &\dec   0.9985(-0.9871,~-1.0014) &\dec  0.9995(-0.9994,-1.0065) &\                              &                              &                               \\
\tableline
% \multicolumn{6}{c}{$\partial S(2\pi)/\partial \lambda$} \\
% n= 1 &                                 &                               &\dec  0.6636(-0.3303,~-0.9008) &\dec  1.0663(-1.3701,~-1.0709) & ~~~~~~~~*~~(~~~~*~~~,~~~~~*~~~) \\
% n= 2 &\dec  0.8219(-0.5941,~-1.0726)   &\dec  0.8869(-0.7089,~-1.2867) &\dec  1.0834(-1.4615,~-1.0229) &\dec  1.0679(-1.3766,~-1.0236) &\dec  0.9733(-0.8722,~-1.0305)  \\
% n= 3 & ~~~~~~~~*~~(~~~*~~~~~,~-1.1164) &\dec  0.9993(-1.0203,~-1.0236) &\dec  0.9925(-0.9866,~-1.0228) &\dec  0.9953(-1.0029,~-1.0439) &\dec  0.9987(-1.0272,~-1.0383)  \\
% n= 4 &\dec  0.9932(-0.9907,~-1.0303)   &\dec  0.9945(-0.9979,~-1.0526) & ~~~~~~~~~*~(~~*~~~~~,~-1.0373)&\dec  0.9993(-1.0316,~-1.0391) &\dec  0.9991(-1.0301,~-1.0358)  \\
% n= 5 & ~~~~~~~~*~~(~~~*~~~~~,~-1.0395) &\dec  0.9983(-1.0231,~-1.0394) &\dec  0.9990(-1.0294,~-1.0416) &\dec  0.9992(-1.0308,~-1.0417) &                                \\
% n= 6 &\dec  0.9991(-1.0300,~-1.0395)   &\dec  0.9999(-1.0399,~-1.0417) &                               &                               &                                \\
% \tableline
\multicolumn{6}{c}{$\Phi_{\rm 1p}$} \\
n= 1 &                             &                             &  ~~~~~~~*~~(~~*~~~~,~0.0118)&\dec  0.2328(0.0004,~0.2419) &   ~~~~~*~~~(~~*~~~~~,~0.3983) \\
n= 2 &   ~~~~~*~~~(~~~*~~~~,~0.2984)&  ~~~~~~~*~~(~~*~~~~,~0.3254)&\dec  0.6947(0.0613,~0.3328) &   ~~~~~*~~~(~~*~~~~,~0.3252)&   ~~~~~*~~~(~~*~~~~~,~0.3231) \\
n= 3 &\dec  0.9773(0.3025,~0.3354) &\dec  1.0250(0.3680,~0.3289) &\dec  1.0065(0.3359,~0.3224) &\dec  0.9901(0.3043,~0.3244) &\dec  0.9935(0.3117,~0.3327) \\
n= 4 &\dec  1.0111(0.3449,~0.3250) &\dec  0.9266(0.1528,~0.3257) &\dec  0.9929(0.3103,~0.3269) &   ~~~~~~*~~(~~*~~~~,~0.3311)&\dec  0.9953(0.3162,~0.3323) \\
n= 5 &\dec  0.9965(0.3188,~0.3215) &\dec  0.9954(0.3164,~0.3218) &\dec  0.9948(0.3148,~0.3187) &\dec  0.9952(0.3158,~0.3568) &                            \\
n= 6 &  ~~~~~~*~~(~~~*~~~~,~0.3215)&\dec  0.9951(0.3154,~0.3225) &                             &                             &                            \\
\tableline
\multicolumn{6}{c}{$\partial \Phi_{\rm 2p}/\partial \lambda $} \\
n= 1 &                               &                               & \dec   -0.2374(~0.0304,~-0.1924) &    ~~~~~~*~~~(~~~*~~~~~,~-1.1986)&  ~~~~~~~*~~~(~~~*~~~~~,~~~~*~~~~) \\
n= 2 &\dec  0.7175(-0.1129,~-0.2521) &\dec  0.2556(-0.0140,~-0.1334) &   ~~~~~~~~~~~~(~~~*~~~~,~-0.7531)&    ~~~~~~*~~~(~~~*~~~~~,~-0.7601)&  ~~~~~~~*~~~(~~~*~~~~~,~~~~*~~~~) \\
n= 3 &\dec  0.4363(-0.0604,~~0.0112) &\dec  0.9908(-0.7323,~-0.7603) & \dec    0.7409(-0.2107,~-0.7529) & \dec   0.6647(-0.1138,~-0.4809) & \dec  0.5793(-0.0453,~-0.6359) \\
n= 4 &\dec  0.8362(-0.3880,~-1.0869) &\dec  0.6399(-0.0869,~-0.6158) & \dec    0.3412(-0.0003,~-0.6030) & \dec   0.6184(-0.0724,~-0.5041) & \dec  0.6071(-0.0630,~-0.1906) \\
n= 5 &\dec  0.4558(-0.0054,~-0.6038) &\dec  0.5885(-0.0485,~-0.6162) & \dec    0.6023(-0.0588,~~~*~~~~) &                                 &                             \\
n= 6 &\dec  0.6035(-0.0599,~-0.4355) \\
\end{tabular}
\end{table}

%=======================================================================
\end{document}